\documentclass[fleqn,usenatbib,useAMS]{mnras}

\usepackage{graphicx}	
\usepackage{amsmath}	
\usepackage{amssymb}	
\usepackage{multicol}        
\usepackage{bm}		
\usepackage{pdflscape}	

\usepackage{etoolbox}
\makeatletter
\patchcmd\@combinedblfloats{\box\@outputbox}{\unvbox\@outputbox}{}{%
   \errmessage{\noexpand\@combinedblfloats could not be patched}%
}%
 \makeatother

\newcommand{\fracb}[2]{\left(\frac{#1}{#2}\right)}

\newcommand{\mean}[1]{\langle{#1}\rangle}

\usepackage[T1]{fontenc}
\usepackage{ae,aecompl}

\usepackage{txfonts}

\title[Afterglow Imaging \& Polarization of Misaligned Structured GRB Jets]{Afterglow Imaging and Polarization of Misaligned Structured GRB Jets and Cocoons: Breaking the Degeneracy in GRB~170817A}

\author[Gill \& Granot 2017]{Ramandeep Gill$^{1,2}$\thanks{Contact e-mail:
\href{mailto:rsgill.rg@gmail.com}{rsgill.rg@gmail.com}}
and Jonathan Granot,$^{1,3}$\thanks{Contact e-mail:
\href{mailto:granot@openu.ac.il}{granot@openu.ac.il}}
\\
$^{1}$Department of Natural Sciences, The Open University of Israel, 
1 University Road, PO Box 808, Raanana 4353701, Israel \\
$^{2}$Physics Department, Ben-Gurion University, P.O.B. 653, 
Beer-Sheva 84105, Israel\\
$^{3}$Department of Physics, The George Washington University, Washington, DC 20052, USA}

\date{Last updated 2018 March 15}

\pubyear{2018}

\begin{document}
\label{firstpage}
\pagerange{\pageref{firstpage}--\pageref{lastpage}}
\maketitle

\begin{abstract}
The X-ray to radio afterglow emission of GRB\;170817A\;/\;GW\;170817  
so far scales as 
$F_\nu\propto\nu^{-0.6}t^{0.8}$ with observed frequency and time, consistent with a single
power-law segment of the synchrotron spectrum from the external shock going into the ambient medium.  
This requires the effective isotropic equivalent afterglow shock energy in the visible region to 
increase as $\sim t^{1.7}$. The two main channels for such an energy increase are (i) \emph{radial}: 
more energy carried by slower material (in the visible region) gradually catches up with the afterglow 
shock and energizes it, and (ii) \emph{angular}: more energy in relativistic outflow moving at 
different angles to our line of sight, whose radiation is initially beamed away from us but its
beaming cone gradually reaches our line of sight as it decelerates. One cannot distinguish between 
these explanations (or combinations of them) using only the X-ray to radio $F_\nu(t)$. Here we 
demonstrate that the most promising way to break this degeneracy is through afterglow imaging and 
polarization, by calculating the predicted evolution of the afterglow image (its size, shape and flux 
centroid) and linear polarization $\Pi(t)$ for different angular and/or radial outflow structures 
that fit $F_\nu(t)$. We consider two angular profiles -- a Gaussian and a narrow core with power-law 
wings in energy per solid angle, as well as a (cocoon motivated) (quasi-) spherical flow with radial 
velocity profile. For a jet viewed off-axis (and a magnetic field produced in the afterglow shock) 
$\Pi(t)$ peaks when the jet's core becomes visible, at $\approx2t_p$ where the lightcurve peaks at $t_p$,
and the image can be elongated with aspect ratios$\;\gtrsim2$. A quasi-spherical flow has an almost circular
image and a much lower $\Pi(t)$ (peaking at $\approx t_p$) and flux centroid displacement $\theta_{\rm fc}$
(a spherical flow has $\Pi(t)=\theta_{\rm fc}=0$ and a perfectly circular image). 
\end{abstract}

\begin{keywords}
gamma-ray burst: short --- stars: neutron --- stars: jets --- polarization --- relativistic processes --- gravitational waves
\end{keywords}




\section{Introduction}
GRB~170817A became the first ever bona fide electromagnetic counterpart \citep[e.g.][and references therein]{Abbott+17b} of a gravitational wave event, GW~170817, detected by advanced 
LIGO/VIRGO observatories, that marked the merger of two neutron stars \citep{Abbott+17a}. A vigorous observation 
campaign that started after this discovery led to the detection of the thermal kilonova emission, that dominated 
the optical and near-infrared energy range at early times, as well as the non-thermal afterglow emission in 
radio and X-rays 
\citep[][]{Abbott+17a,Abbott+17b,Abbott+17c,Alexander+17,Chornock+17,Coulter+17,Covino+17,Cowperthwaite+17,Drout+17,Goldstein+17,Haggard+17,Hallinan+17,Kasliwal+17,Lyman+18,Margutti+17,Mooley+18,Nicholl+17,Pian+17,Ruan+18,Smartt+17,SS17,Tanvir+17,Troja+17,Valenti+17,Villar+17}.
The broadband afterglow emission from the short-hard gamma-ray burst GRB~170817A, which has been regularly 
monitored in radio, optical, and X-rays 
therefore presented a golden opportunity to improve our understanding 
of the properties of relativistic outflows in GRBs, and in particular their geometry and how their energy 
is distributed as a function of angle and proper velocity. The afterglow emission \textcolor{black}{continued to rise in 
flux until $\gtrsim115$~days post-merger \citep[e.g.][]{Lyman+18,Margutti+18,Mooley+18,Ruan+18,Troja+18}, where 
it might have shown a plateau in the light-curve at $\sim 138$~days \citep{D'Avanzo+18,Resmi+18} and a peak in the X-ray \citep{Margutti+18} 
and radio \citep{Dobie+18} lightcurves at $\sim 150 - 160$~days.} It was the rising flux that seriously challenged the simple model 
of a narrowly beamed, sharp-edged, ultra-relativistic homogeneous jet.

The leading types of models that have been successful at explaining the rising afterglow flux thus far feature an outflow structure that is predominantly either (i) \emph{radial}: a broad distribution of energy with proper velocity $u=\Gamma\beta$ in the outflow with more energy carried by slower material (in the visible region)
that gradually catches up with the afterglow shock and energizes it 
\citep[with a wide-angle quasi-spherical mildly relativistic flow; e.g.][]{Kasliwal+17,D'Avanzo+18,FV18,GNP18,Hotokezaka+18,Mooley+18,NP18,Troja+18}, 
or (ii) \emph{angular}: a jet with angular structure 
containing an energetic and initially highly relativistic 
core and sharply falling lower energy wings along which our line of sight is located 
\citep[e.g.][]{LK17b,Lazzati+17a,Lazzati+17b,Troja+17,D'Avanzo+18,Margutti+18,Resmi+18,Troja+18}. 
In the latter explanation the radiation from 
the energetic parts of the jet near its core is initially beamed away from us, and 
gradually becomes visible as the jet decelerates by sweeping up the external medium. 
In order to better distinguish between such types of models, or combinations of them, 
it is useful to look at where most of the energy resides and when it contributes to 
the observed emission, i.e. when it decelerates for a radial structure, or when its 
beaming cone reaches our line of sight for a jet with angular structure. Both scenarios 
can fit the radio and X-ray observations and yield similar late-time behavior of the lightcurves. 
To break this degeneracy in the two models other diagnostics must be considered.

In this work, we demonstrate that the most promising way to unveil the properties of the outflow, and the 
distribution of its energy with angle and/or proper velocity $u$, is through 
afterglow imaging and polarization. To this end, we consider different physically motivated angular and radial outflow 
structures that can fit the observed lightcurves and spectrum, $F_\nu(t)$, and calculate for them the predicted evolution of
the afterglow image -- size, shape, and flux centroid -- and linear polarization. The paper is structured as follows. 
We start by describing the dynamics and structure of the different outflow profiles that are considered here in \S2. The lateral dynamics are ignored, and the possible implications are discussed in \S5. Next, 
in \S3, we assume that the underlying afterglow emission mechanism is synchrotron and calculate lightcurves for off-axis emission 
for the different models, which we also compare with radio, optical, and X-ray observations. We further assume 
that the magnetic field in the shocked ejecta is completely tangled in the plane orthogonal to the shock normal 
and calculate the degree of linear 
polarization for all the models in \S4. In \S5, we show the radio images for the different models and calculate the 
temporal evolution of important characteristics, such as the flux centroid, mean image size, and its axial ratio. Finally, 
in \S6, we discuss the importance and feasibility of the diagnostics that are presented in this work and that hold the 
potential to break the degeneracy between structured jets and quasi-spherical outflows.

\section{The Outflow Structure and Dynamics}

\begin{figure}
    \centering
    \includegraphics[width=0.48\textwidth]{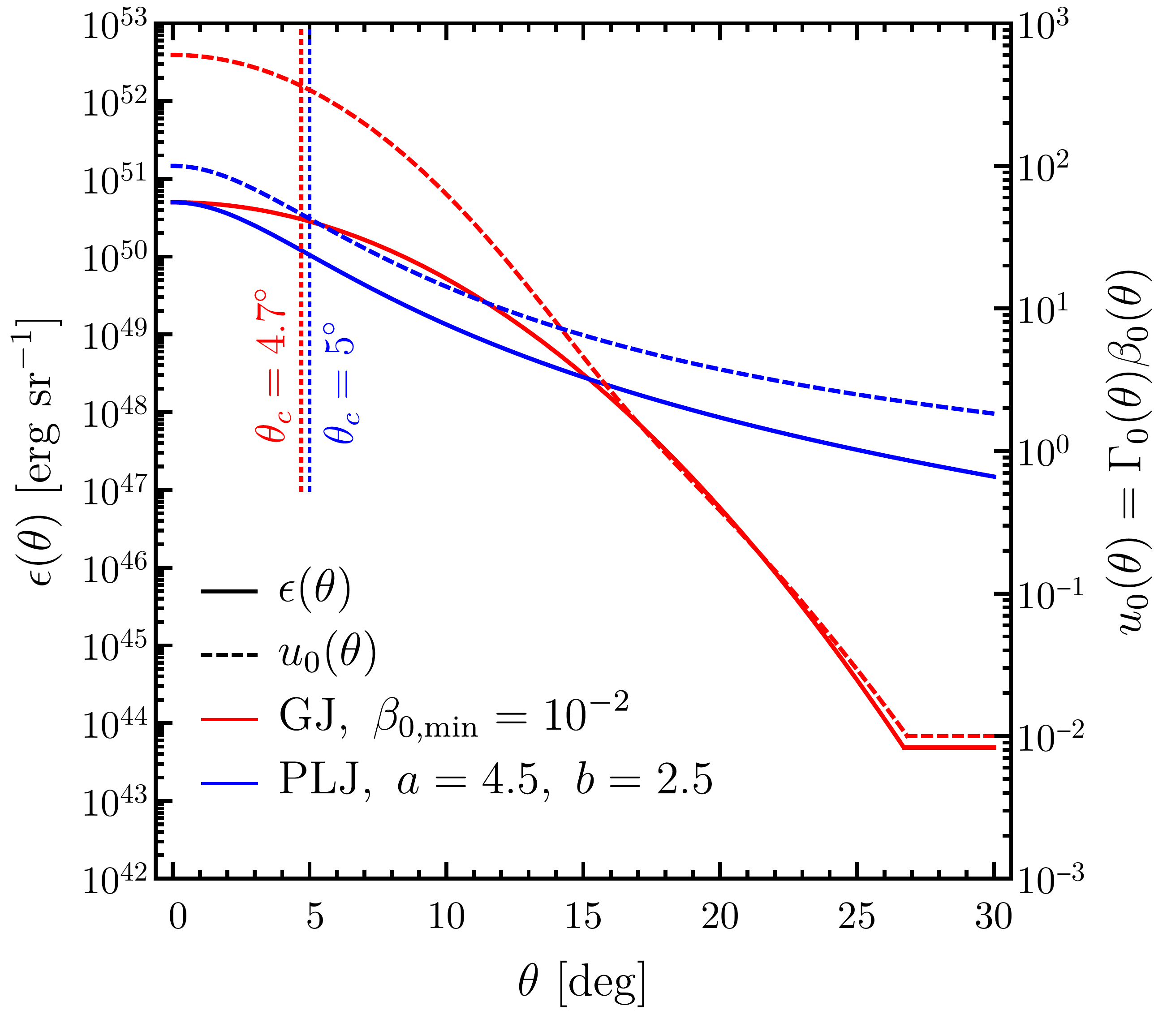}
    \caption{Angular profile of the energy per solid angle $\epsilon(\theta)=E_{\rm k,iso}(\theta)/4\pi$ and initial proper 
    velocity $u_0(\theta)=\Gamma_0(\theta)\beta_0(\theta)$ for the two structured jet models considered 
    here -- (GJ) Gaussian jet and (PLJ) power-law jet. The core angle $\theta_c$ beyond which $\epsilon$ 
    and $u_0$ start to drop sharply is shown with the vertical dotted line.}
    \label{fig:jet-profile}
\end{figure}

\subsection{A Thin Shell with Local Spherical Dynamics}
For simplicity we restrict the treatment in this work to axi-symmetric outflows. For clarity, 
let us define a structured jet or outflow as one in which the energy per 
unit solid angle $dE/d\Omega\equiv\epsilon(\theta)$ and/or the Lorentz factor (LF) $\Gamma(\theta,r)$ 
of the jet vary smoothly with the angle $\theta$ from the jet symmetry axis \citep[e.g.][]{MRW98}. As the jet expands 
into the external medium, it sweeps up mass $dm(r)= \rho(r)4\pi r^2dr$, where 
$\rho(r)=n(r)m_p=A r^{-k}$ (where $m_p$ is the proton mass) and $n(r)$ are the external mass density 
and number density, respectively, which are assumed here to have a power-law profile with radius $r$. 
For short GRBs that explode in the interstellar medium (ISM) of their host galaxy one expects a uniform 
density ($k=0$). For long GRBs the outflow expands into a density profile produced by the stellar wind 
of their massive star progenitor, for which $k=2$ may be expected for a steady wind.\footnote{Or for a 
wind with a constant wind mass loss rate to velocity ratio, $\dot{M}_w/v_w$. Other values of $k$ 
or a non power-law profile are possible if $\dot{M}_w$ and/or $v_w$ vary in the last stages of the massive star's life \citep[e.g.][]{G-S96,CL00,R-R01,CLF04,R-R05,vanMarle06,Kouveliotou13}.} 

A thin shell approximation is used for the layer of shocked external medium that carries most of the 
energy and dominates the observed emission. The lateral dynamics
are ignored in this simple treatment, and instead the dynamics at each angle $\theta$ are assumed to 
be independent of other angles. The local dynamics at each $\theta$ are assumed to correspond to a 
spherical flow with the local isotropic equivalent jet energy $E_{\rm k,iso}(\theta)=4\pi\epsilon(\theta)$. 
At an early stage the shell is assumed to coast with a bulk LF $\Gamma_0(\theta)$ until the deceleration 
radius $r_d(\theta)$, where most of its energy is used up to  accelerate the shocked external medium 
to $u\approx u_0(\theta)$ and heat it up to a similar thermal proper velocity, so that 
$m[r_d(\theta)]u_0^2(\theta)c^2=E_{\rm k,iso}(\theta)$. Here $m(r)=[4\pi A/(3-k)]r^{3-k}$ is the isotropic 
equivalent swept-up rest mass up to radius $r$, and $u_0(\theta)=\Gamma_0\beta_0=[\Gamma_0^2(\theta)-1]^{1/2}$ 
is the dimensionless proper velocity, where $u_0\approx\Gamma_0$ for $\Gamma_0\gg1$ and $u_0\approx\beta_0$ for 
$\Gamma_0-1\ll1$. The deceleration radius is given by
\begin{eqnarray}
    r_d(\theta) &=& \left[\frac{(3-k)E_{\rm k,iso}(\theta)}{4\pi Ac^2u_0^2(\theta)}\right]^{1/(3-k)} \\
    &\approx& 1.3\times10^{17}E_{53}^{1/3}u_{0,2}^{-2/3}n_0^{-1/3}~{\rm cm}\quad(k=0)\ , \nonumber
\end{eqnarray}
where $Q_x$ is the quantity $Q$ in units of $10^x$ times its c.g.s units. Beyond this radius the shell 
starts to decelerate as it continues to sweep up more mass and its evolution becomes self-similar, such 
that $u(\theta,r)\propto r ^{(k-3)/2}$ both during the relativistic phase \citep{BM76},
and during the Newtonian Sedov-Taylor phase. Radiative losses are neglected, and an adiabatic evolution is assumed from coasting phase through the relativistic and Newtonian self-similar phases. This can be reasonably described as follows. The original shell of rest mass $m_0$ and initial energy $E_0=(\Gamma_0-1)m_0c^2$ is assumed to remain cold as it decelerates and have a kinetic energy of $(\Gamma-1)m_0c^2$. The swept-up external medium of rest mass $m(r)$ has similar bulk and thermal proper velocities of $u$, so that its total energy excluding its rest energy is $m(r)c^2u^2 = m(r)c^2(\Gamma^2-1)$. Therefore, energy conservation reads
$(\Gamma_0-1)m_0=E_0/c^2=m_0(\Gamma-1)+m(r)(\Gamma^2-1)$. Defining the dimentionless radius $\xi(\theta)\equiv r/r_d(\theta)$ one obtains that $m/m_0 = \xi^{3-k}/(\Gamma_0+1)$, and energy conservation reads \citep{PK00}
\begin{equation}\label{eq:energy1}
    \frac{\xi^{3-k}}{\Gamma_0+1}(\Gamma^2-1)+\Gamma-\Gamma_0=0\ ,
\end{equation}
with the solution
\begin{equation}\label{eq:Gamma}
    \Gamma(\xi) = \frac{\Gamma_0+1}{2}\xi^{k-3}
    \left[\sqrt{1+\frac{4\Gamma_0}{\Gamma_0+1}\xi^{3-k}+\fracb{2\xi^{3-k}}{\Gamma_0+1}^2}-1\right]\ .
\end{equation}
The expression for $\Gamma(\xi)$ 
presented above is quite general and applies both when $\Gamma_0$ is ultrarelativistic as well as 
when $\Gamma_0\gtrsim1$. It is similar to the expression presented in equation~(4) of \citet{PK00} 
in the limit $\Gamma_0\gg1$.

\subsection{Structured Jets -- with an Angular Profile}

In this work, we consider two distinct angular profiles for the structured jet: (i) A \textit{Gaussian jet} (GJ)
for which both $\epsilon(\theta)$ and $\Gamma_0(\theta)-1$ have a Gaussian profile with a standard deviation or 
core angle $\theta_c$ \citep[e.g.][]{ZM02,KG03,Rossi+04},
\begin{equation}
    \frac{\epsilon(\theta)}{\epsilon_c}=
    \frac{\Gamma_0(\theta)-1}{\Gamma_c-1} = 
    \max\left[\exp\left(-\frac{\theta^2}{2\theta_c^2}\right)~,~\exp\left(-\frac{\theta_*^2}{2\theta_c^2}\right)\right]~,
\end{equation}
where $\epsilon_c$ and $\Gamma_c$ are the core energy per unit solid angle and initial core LF, 
with a floor at $\theta>\theta_*$ corresponding to $\beta_{0,\rm min} = 0.01$, and 
(ii) a \textit{power-law jet} (PLJ) for which $\epsilon(\theta)$ and $\Gamma_0(\theta)-1$ decrease as a 
power-law in $\theta$ outside of the core angle, $\theta_c$, such that \citep[e.g.][]{Rossi+02,Rossi+04,GK03,KG03}
\begin{eqnarray}
    &&\epsilon(\theta) = \epsilon_c\Theta^{-a}\quad,\quad\Theta=\sqrt{1+\fracb{\theta}{\theta_c}^2}\ , \\
    &&\Gamma_0(\theta) = 1+(\Gamma_c-1)\Theta^{-b}\ .
\end{eqnarray}
Fig.~\ref{fig:jet-profile} shows the two jet angular profiles for our selected parameters that provide a 
good fit to the afterglow radio to X-ray lightcurves.

\begin{figure}
    \centering
    \includegraphics[width=0.48\textwidth]{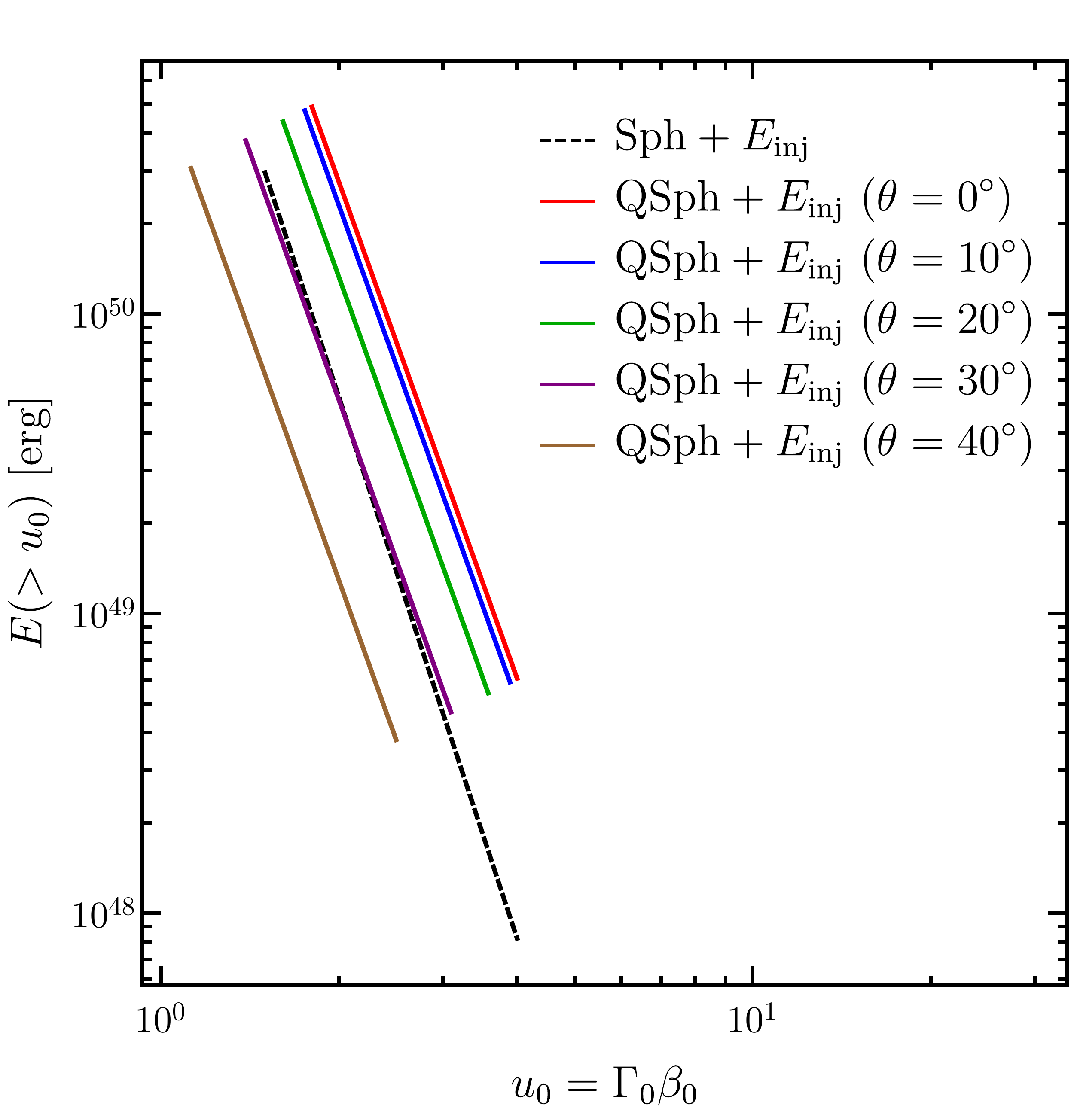}
    \includegraphics[width=0.48\textwidth]{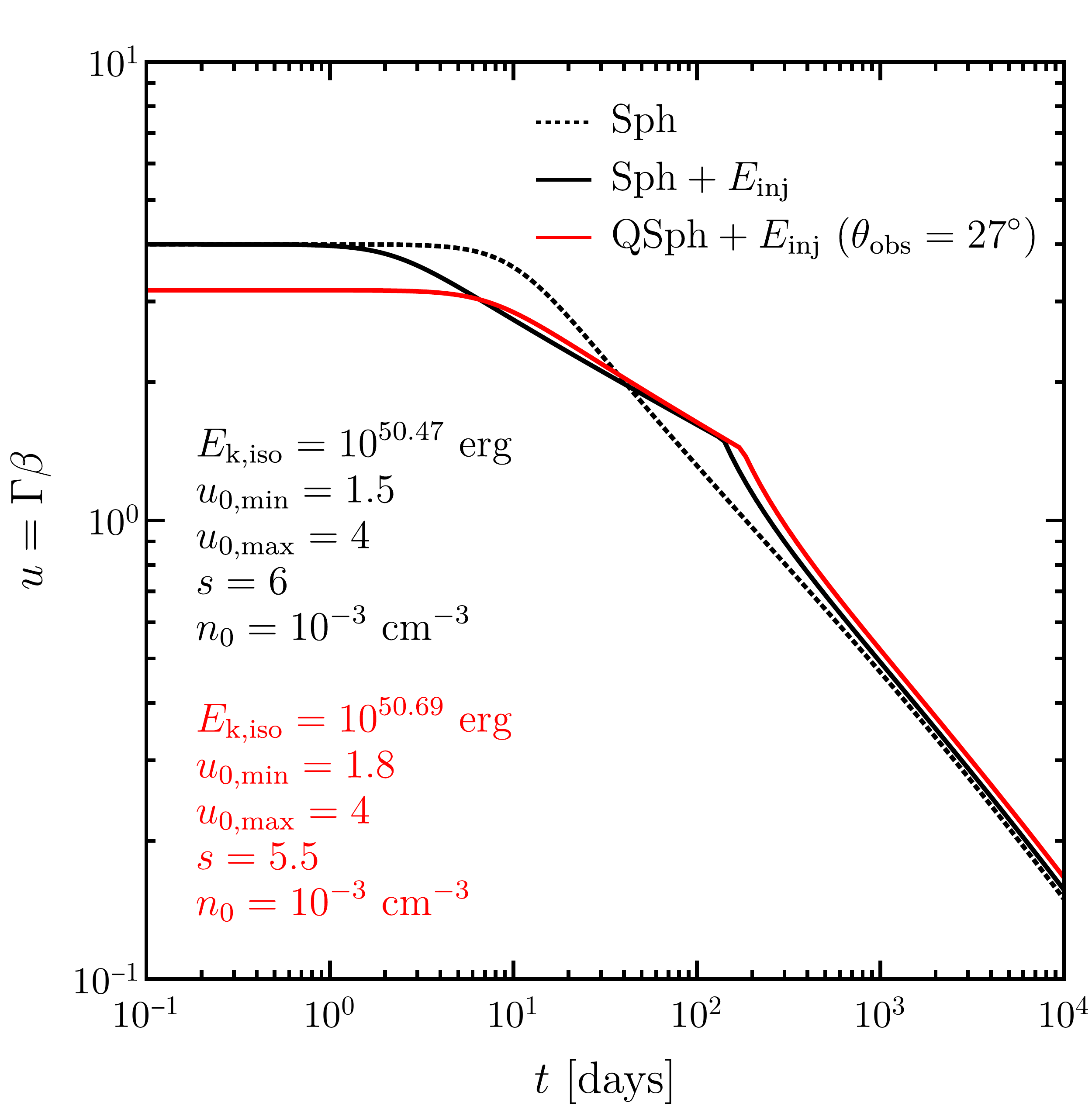}
    \caption{(Top) Initial radial velocity stratification for the spherical shell (Sph) 
    and quasi-spherical shell (QSph) models. The relevant parameters are shown in the 
    bottom panel. (Bottom) Dynamical evolution of the Sph model 
    with and without energy injection (due to initial velocity stratification). Also shown 
    is the case of QSph model with energy injection for a viewing angle of $\theta_{\rm obs}=27^\circ$.}
    \label{fig:uTz}
\end{figure}

\subsection{Outflows with a Radial Profile, $E_{\rm k,iso}(u)$: (Quasi-)Spherical Shell with Energy Injection}
If the jet cannot break out of the dynamical ejecta of the binary neutron star (BNS) merger and/or 
the neutrino-driven wind that is launched just after the merger, then it will be chocked. In this 
case all of the jet's energy is transferred to a cocoon consisting of shocked jet material and 
shocked surrounding material, where the latter quickly becomes dominant energetically. The cocoon 
ultimately breaks out of the surrounding medium that was ejected during the merger and can reach 
mildly relativistic velocities (typically $\Gamma$ of up to a few or several). The emerging cocoon 
is expected to form a wide-angle, quasi-spherical flow, and if the external medium's density sharply 
drops near its outer edge then the cocoon-driven shock would accelerate as it propagates down 
that density gradient and form an asymptotic distribution of energy with proper velocity $u$ in 
the resulting outflow that sharply drops with $u$. This is the `cocoon' scenario \citep[e.g.][]{Kasliwal+17,Mooley+18,NP18} 
that has been suggested to explain the initial sub-luminous gamma-ray and the later broadband afterglow 
emission from GRB~170817A. 

Alternatively, the BNS merger can give rise to a dynamical 
ejecta driven by the shock wave that is formed as the two NSs collide that crosses the stars and accelerates down the sharp density gradient in their outer layers, and form an energy distribution that drops less sharply with $u$, $E(>u)\propto u^{-1.1}$ for $u\gg 1$ \citep[e.g.][]{KIS14}. 

In both cases  most of the energy resides in the slower moving material as compared to a faster moving head of the ejecta. The fastest moving ejecta sweeps up the external medium by driving a relativistic forward afterglow shock into it, and is itself decelerated by a reverse shock, where the two shocked regions are separated by a contact discontinuity \citep[e.g.][]{SP95}. As more external medium is constantly swept up by the forward shock, this double-shock structure gradually decelerates, allowing slower and more energetic ejecta to catch up with it and energize it \citep[e.g.][]{SM00,NS06}. This energy injection by the slower and more energetic ejecta results in a slower deceleration of the afterglow shock compared to the case of no energy injection. If $E(>u)$ falls sharply enough with $u$ resulting in a sufficiently fast energy injection rate, this can lead to a gradual rise in the observed flux (see Appendix~\ref{sec:App-radial}). 

The distribution of the ejecta's energy with its initial proper velocity $u_0=u(t_0)$ can be parameterized as a power-law \citep[e.g.][]{Mooley+18}, such that 
\begin{equation}
 E(>u_0) = E_0\fracb{u_0}{u_{0,\rm max}}^{-s}\quad{\rm for }\quad u_{0,\rm min} \leq u_0 \leq u_{0,\rm max}~.
\end{equation}
Here $E_0=(\Gamma_0-1)m_0c^2$ is the energy in the fastest ejecta, of rest mass $m_0$, that is assumed to be cold and initially coasting at $\Gamma_0 = (1+u_{\rm0,max}^2)^{1/2}$. It is related to the total isotropic equivalent kinetic energy through
\begin{equation}
E_{\rm k,iso}= E_0\fracb{u_{\rm0,max}}{u_{0,\rm min}}^{s}\ .
\end{equation}
The observed flux rise suggests a steep distribution with $s\sim5-6$ \citep[see Appendix~\ref{sec:App-radial} or][]{Mooley+18,NP18}, which is too steep for the dynamical ejecta scenario mentioned above.

In this scenario of radial gradual energy injection by slower ejecta the deceleration radius is given by
\begin{eqnarray}
    r_d(\theta) &=& \left[\frac{(3-k)E_0(\theta)}{4\pi Ac^2u_{\rm0,max}^2(\theta)}\right]^{1/(3-k)} \\
    &\approx& 1.3\times10^{16}E_{0,50}^{1/3}u_{\rm0,max,2}^{-2/3}n_0^{-1/3}~{\rm cm}\quad(k=0,~\theta=0)\ . \nonumber
\end{eqnarray}
The dynamical evolution of the emitting region, $u(\xi)$, can be obtained by (numerically) solving the relevant generalization of Eq.~(\ref{eq:energy1}) -- the dimensionless energy equation,
\begin{equation}\label{eq:energy2}
    \frac{\xi^{3-k}}{\Gamma_0+1}u^2+\sqrt{1+u^2}-\Gamma_0\min\left[\fracb{u_{\rm0,max}}{u}^{s},\fracb{u_{\rm0,max}}{u_{\rm0,min}}^{s}\right] = 0\ .
\end{equation} 

We consider two angular profiles for such a wide-angle flow: (i) a uniform spherical shell, for which 
\begin{equation}
\displaystyle
\epsilon(\theta)=\frac{E_{\rm k,iso}}{4\pi}= \frac{E_0}{4\pi}\fracb{u_{\rm0,max}}{u_{0,\rm min}}^{s}~,
\end{equation}
and (ii) a quasi-spherical angular profile that is given by
\begin{equation}
    \frac{\epsilon(\theta)}{\epsilon_0} =
    \frac{u_{0,\rm min}(\theta)}{u_{\rm min,0}}=
    \frac{u_{0,\rm max}(\theta)}{u_{\rm max,0}}=
    \frac{\zeta+\cos^2\theta}{\zeta+1}\ ,
\end{equation}
with $\zeta=0.1$. The parameter $\zeta$ is chosen to mimic a floor at 
$u_{0,\{\rm min,max\}}(\theta=\pi/2)$. 
Fig.~\ref{fig:uTz} shows the outflow energy distributions with proper 
velocity, $E_{\rm k,iso}(>u_0)$ ({\it top panel}), and the evolution of $u$ 
with the observed time $t$ ({\it bottom panel}), for these two outflow profiles. 
The plots are shown for our selected parameters that provide a good 
fit to the afterglow radio to X-ray lightcurves.


\begin{figure}
\centering
\includegraphics[width=0.3\textwidth]{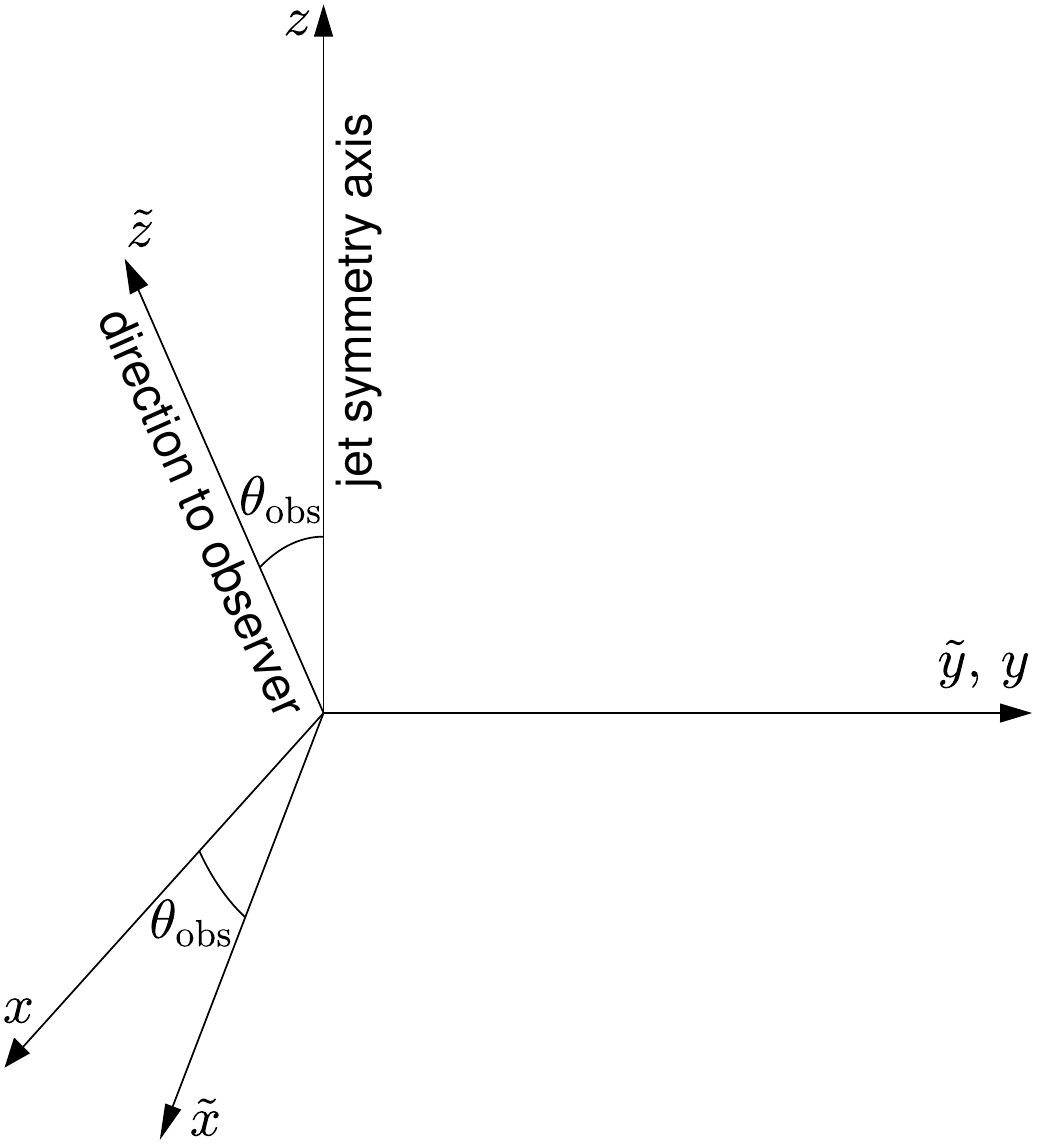}
\caption{Coordinate system used to calculate the observed afterglow flux density and image. The $z$-axis is the outflow's symmetry axis, while the $\tilde{z}$-axis points to the observer and is in the $x$-$z$ plane at an angle of $\theta_{\rm obs}$ 
from the $z$-axis. The $y$ and $\tilde{y}$ axes coincide. The afterglow image is in the plane of the sky, i.e. in the $\tilde{x}$-$\tilde{y}$ plane.}
\label{fig:coord}
\end{figure}



\section{Calculating the Observed Radiation}

\subsection{Synchrotron Emission from the Forward Shock}
Synchrotron radiation is usually the dominant emission mechanism throughout the afterglow. 
Accordingly, we consider synchrotron emission from shock accelerated electrons within 
the shocked external medium behind the forward shock. For simplicity we ignore the effects 
of synchrotron self-absorption and inverse Compton scattering. They are not expected to be very important for our purposes.\footnote{Synchrotron-self Compton (SSC) can increase the cooling of the synchrotron emitting relativistic electron, and reduce their cooling break frequency $\nu_c$ by a factor of $(1+Y)^2$ where $Y$ in the Compton parameter. However, we have verified that this effect does not significantly affect our tentative fits to the data, as $\nu_c$ still remains (at least marginally) above the measured X-ray energy range.} We do not consider here the emission from the long-lived reverse shock
\citep[see e.g.][]{SM00}, but in the relevant power-law segment of the synchrotron spectrum its emission is expected to be sub-dominant compared to that of the forward shock for an electron energy distribution power law index $p>2$ (where in our case $p\approx2.2$ is inferred from observations).

The proper internal energy 
density of the postshock layer is $e'=(\Gamma-1)n'm_pc^2$, where $n'\approx4\Gamma n(r)$ is the proper electron number 
density. A fraction $\epsilon_e$ of this energy is shared by the relativistic electrons that have a mean LF of 
$\langle\gamma_e\rangle=\epsilon_e(m_p/m_e)(\Gamma-1)$ and are shock accelerated to form a power-law distribution, 
such that $n'(\gamma_e)\propto\gamma_e^{-p}$ for $\gamma_m\leq\gamma_e\leq\gamma_M$, with 
$\gamma_m = [(p-1)/(p-2)]\langle\gamma_e\rangle$ for $p>2$.

In the rest frame of the emitting plasma, the local comoving sychrotron emissivity (power per unit frequency per unit volume) can be expressed as a broken power-law:
\begin{eqnarray}
    \frac{P'_{\nu'}}{P'_{\nu',{\rm max}}} = \left\{
    \begin{array}{lc}
        (\nu'/\nu_m')^{1/3} & \nu' < \nu_m' < \nu_c'\\
        (\nu'/\nu_c')^{1/3} & \nu' < \nu_c' < \nu_m' \\
        (\nu'/\nu_m')^{(1-p)/2} & \nu_m' < \nu' < \nu_c' \\
        (\nu'/\nu_c')^{-1/2} & \nu_c' < \nu' < \nu_m' \\
        (\nu'/\nu_m')^{(1-p)/2}(\nu'/\nu_c')^{-1/2} & \nu' > {\rm max}(\nu_m',\nu_c')
    \end{array}\right.
\end{eqnarray}
The flux normalization and break frequencies are
\begin{eqnarray}
    P'_{\nu',{\rm max}} &=& 0.88\frac{512\sqrt{2\pi}}{27}\fracb{p-1}{3p-1}
    \frac{q_e^3}{m_ec^2}(\epsilon_Be')^{1/2}n'~, \\
    \nu_m' &=& \frac{3\sqrt{2\pi}}{8}\fracb{p-2}{p-1}^2\frac{q_e}{m_e^3c^5}
    \epsilon_B^{1/2}\epsilon_e^2(e')^{5/2}(n')^{-2}~, \\
    \nu_c' &=& \frac{27\sqrt{2\pi}}{128}\frac{q_em_ec}{\sigma_T^2}(\epsilon_Be')^{-3/2}
    \fracb{\Gamma}{t_{\rm lab}}^2~,
\end{eqnarray}
where $\nu_m'$ and $\nu_c'$ are, respectivley, the typical synchrotron frequencies, expressed in the comoving 
frame, corresponding to electrons moving with Lorentz factors $\gamma_m$ and $\gamma_c$, where the 
latter are cooling at the dynamical time. Also, in the above equations $q_e$ is the elementary charge 
and $\sigma_T$ is the Thomson cross-section. The swept-up external rest mass per unit shock 
area is $m(r)/4\pi r^2 = Ar^{1-k}/(3-k)$, which for a uniform shell implies a comoving radial 
width of $\Delta'=r/4(3-k)\Gamma$. The shell's isotropic equivalent comoving spectral 
luminosity $L'_{\nu'}$ (the total power per unit frequency assuming a spherical shell with 
the local properties at any given angle $\theta$ from the jet axis) is related to $P'_{\nu'}$ 
through the volume of the emitting region, and is therefore given by 
$L'_{\nu'}/P'_{\nu'}=L'_{\nu',{\rm max}}/P'_{\nu',{\rm max}}=V'=4\pi r^2\Delta'=\pi r^3/(3-k)\Gamma\propto r^3/\Gamma(r)$.

The synchrotron emissivity given above implicitly assumes that all electrons in the emission 
region contribute towards the afterglow emission. That may not be true and only a fraction 
$\xi_e$ of the total number of electrons may actually be shock accelerated into a power-law 
distribution to produce the observed synchrotron emission. In that case, a simple parameterization 
of $E\to E/\xi_e$, $n\to n/\xi_e$, $\epsilon_e\to\epsilon_e\xi_e$, and $\epsilon_B\to\epsilon_B\xi_e$ 
for $m_e/m_p<\xi_e<1$ would yield the same spectral flux $F_\nu$ \citep{EW05}. In this work, we 
assume $\xi_e=1$.

\begin{figure*}
    \centering
    \includegraphics[width=0.45\textwidth]{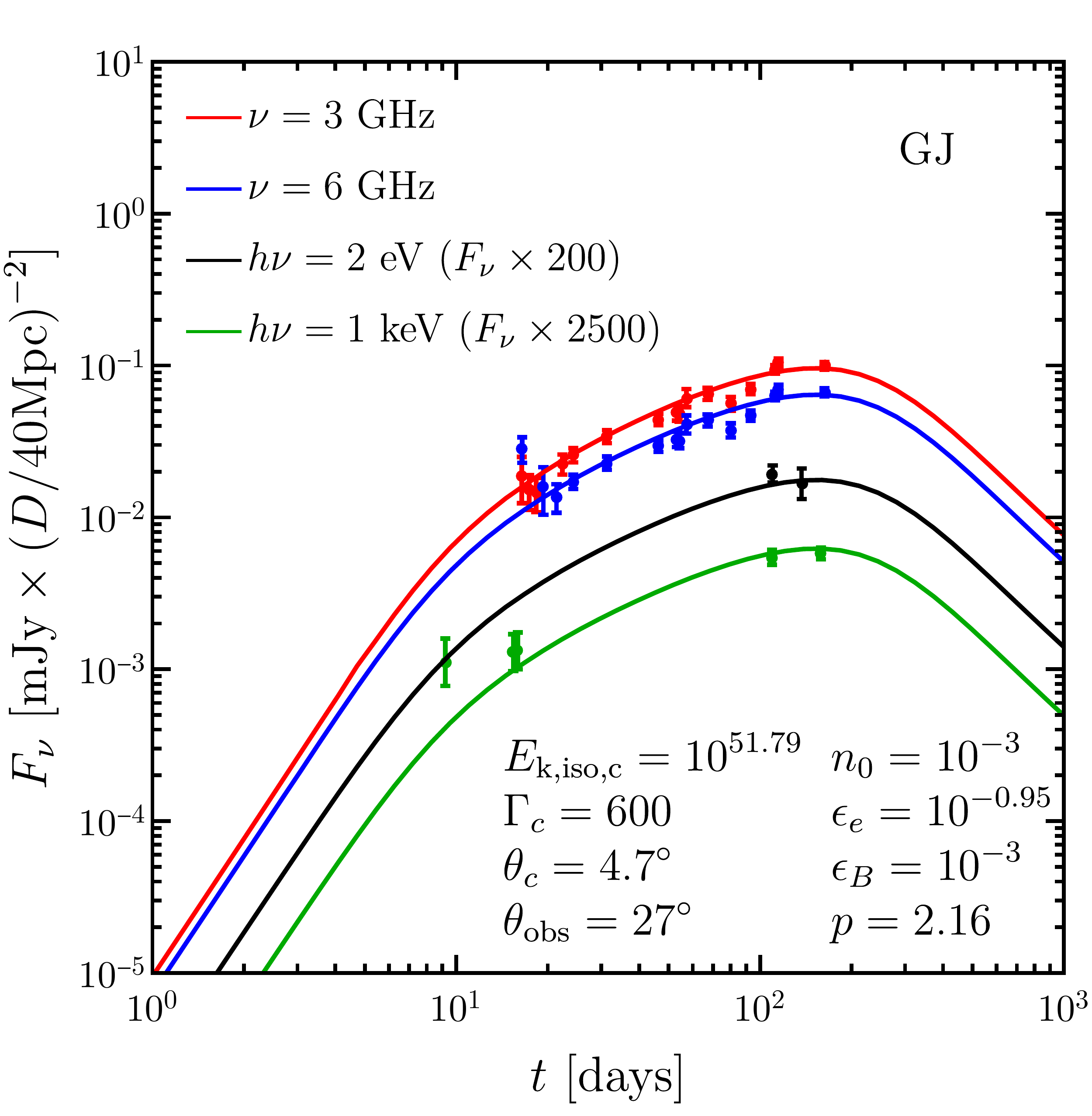}\hspace{1cm}
    \includegraphics[width=0.45\textwidth]{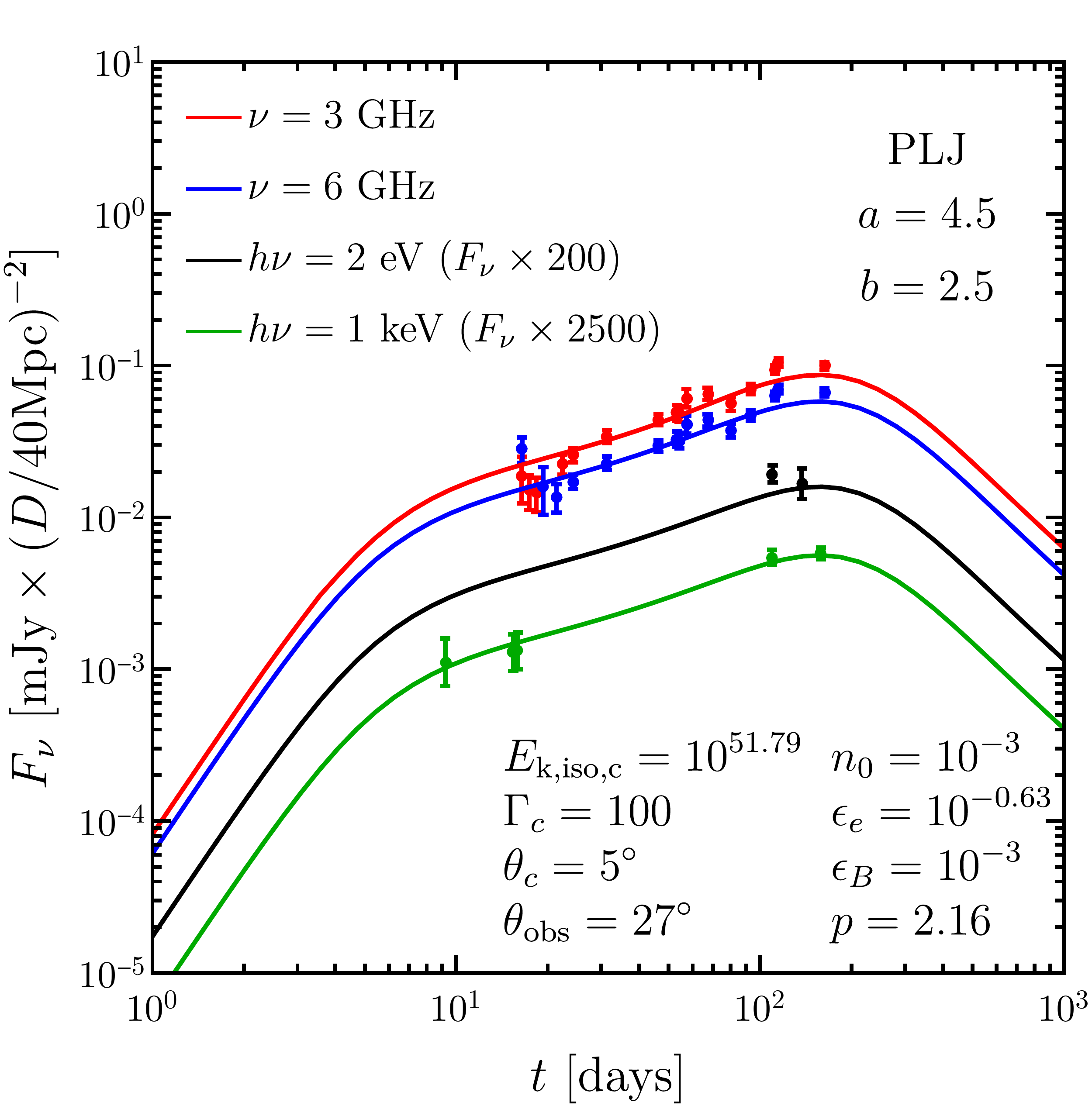}
    \includegraphics[width=0.45\textwidth]{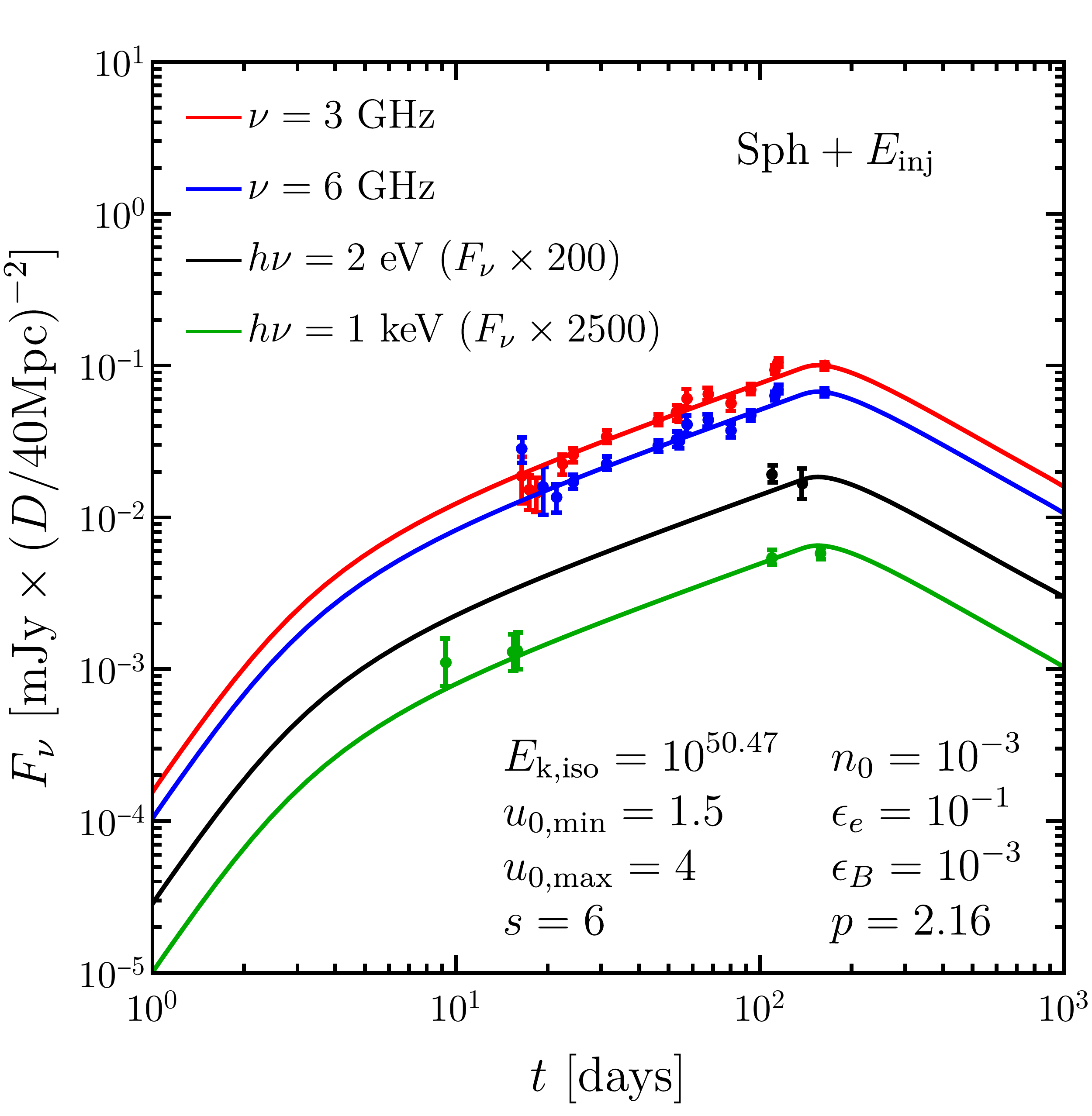}\hspace{1cm}
    \includegraphics[width=0.45\textwidth]{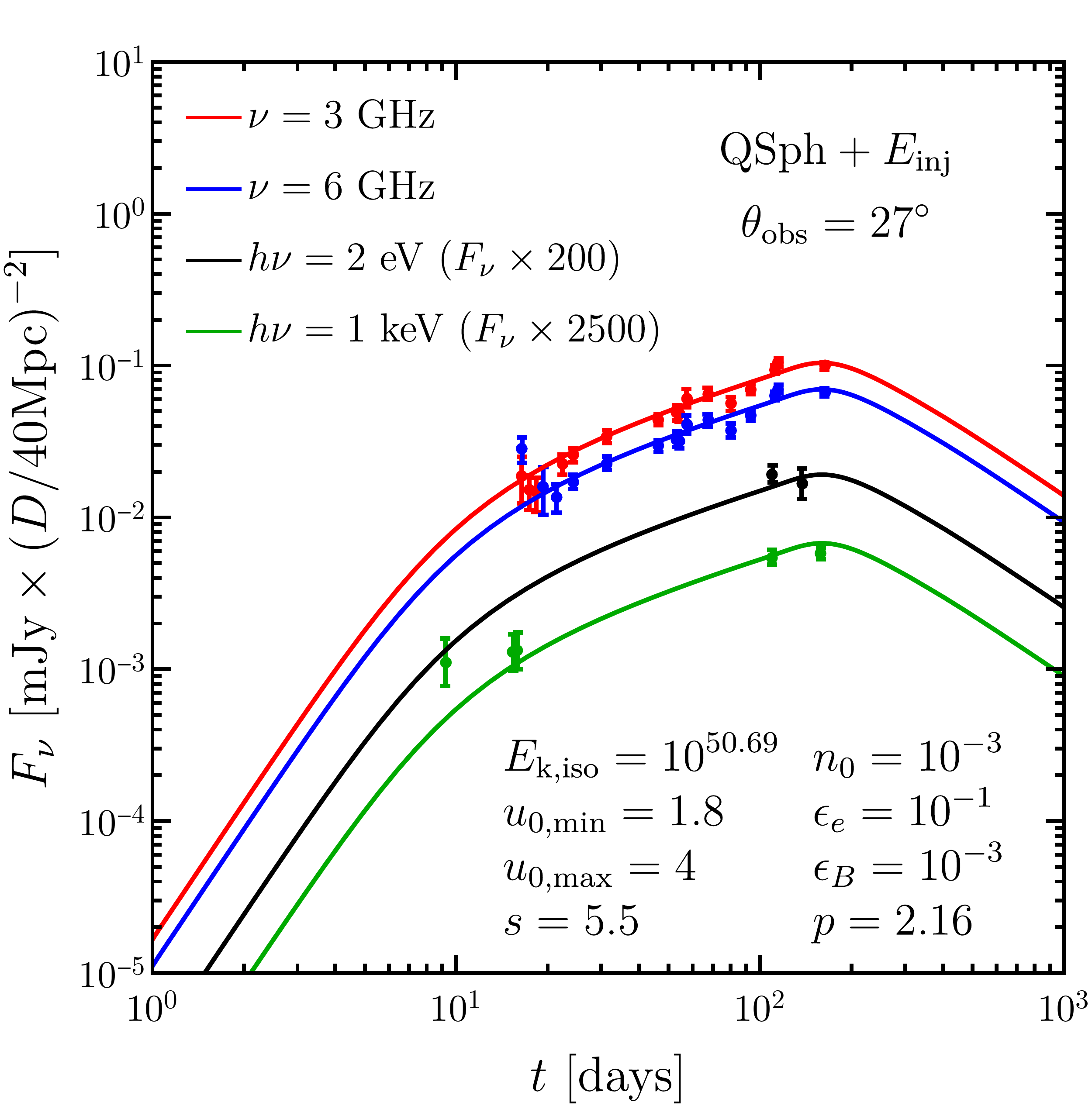}
    \caption{Comparison of radio, optical, and X-ray lightcurves for the gaussian jet (GJ; top-left), 
    power-law jet (PLJ; top-right), spherical shell with energy injection (Sph; bottom-left), 
    and quasi-spherical shell with energy injection (QSph; bottom-right) to the afterglow data 
    for GRB 170817A.}
    \label{fig:lc}
\end{figure*}

\begin{figure}
    \centering
    \includegraphics[width=0.48\textwidth]{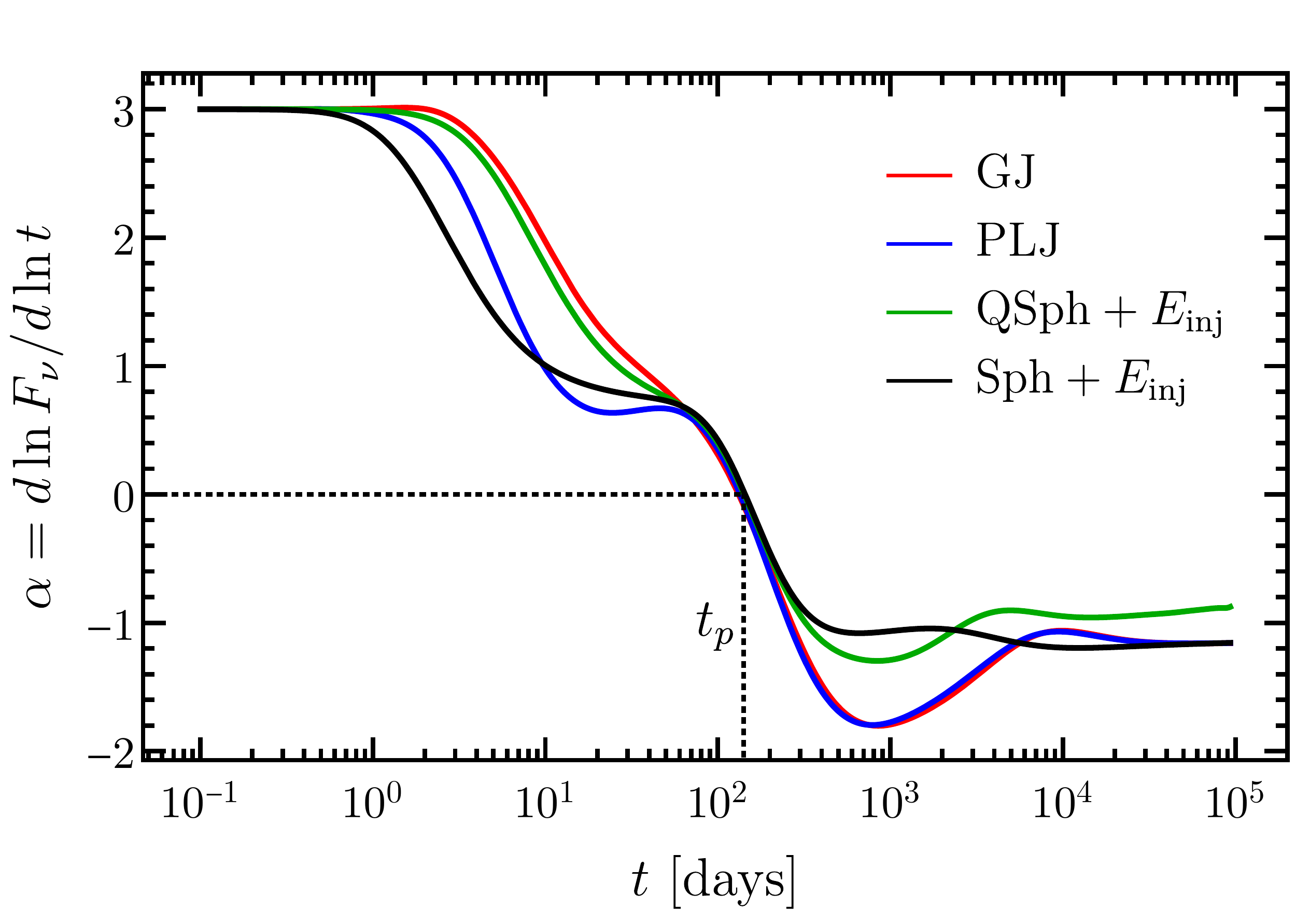}
    \caption{Temporal index of the lightcurves shown for the GJ, PLJ, QSph, and Sph models.}
    \label{fig:fnu-slope}
\end{figure}

\subsection{Observer frame spectrum}
The emission originates from a shocked layer 
of lab-frame width $\Delta=\Delta'/\Gamma\approx r/4(3-k)\Gamma^2$ and from polar angles $0\leq\theta\leq\theta_j$, 
where $\theta$ is measured from the jet axis and $\theta_j$ represents the jet's semi-aperture. 
However, here we make a simplifying assumption and ignore the radial structure of the emitting volume and instead 
consider an infinitely thin-shell. This thin-shell is located at a normalized radial distance $\xi$ from the central source at 
the lab-frame time
\begin{equation}
    t_{\rm lab} = \frac{r_d}{c}\int_0^{\xi}\frac{d\xi'}{\beta(\xi')}~,
\end{equation}
which depends on the dynamics through $\beta=\sqrt{1-\Gamma^{-2}}\approx1-1/2\Gamma^2$ for $\Gamma\gg1$. 
The direction to the observer, $\hat n$, is at an angle $\theta_{\rm obs} = \cos^{-1}(\hat n\cdot \hat z)$ 
from the jet axis, which we conveniently choose to point in the $\hat z$ direction 
(see Figure \ref{fig:coord}). The arrival time $t$ to a distant observer of a photon 
emitted at radius $r$ and angle $\tilde\theta = \cos^{-1}(\hat r\cdot\hat n)$ 
from the LOS, from a source located at a redshift $z$ corresponding to a luminosity 
distance $d_L(z)$, is given by
\begin{equation}\label{eq:tobs}
    t_z\equiv\frac{t}{(1+z)} = t_{\rm lab}-\frac{r\tilde\mu}{c}~,
\end{equation}
where $\tilde\mu=\cos\tilde\theta=\hat{r}\cdot\hat{n}$. When the observer is exactly 
along the jet's axis,  $\hat{n}=\hat{z}$, $\theta_{\rm obs}=0$ and $\tilde{\theta}=\theta$. 

The spectral flux can be expressed using the isotropic comoving spectral 
luminosity $L'_{\nu'}$ such that \citep[e.g.][]{Granot05}
\begin{equation}\label{eq:Fnu}
    F_\nu(t) = \frac{(1+z)}{16\pi^2d_L^2}\int\tilde\delta_D^3L'_{\nu'}d\tilde\Omega~,
\end{equation}
where $\tilde\delta_D = [\Gamma(1-\beta\tilde\mu)]^{-1}\approx2\gamma/(1+\Gamma^2\tilde\theta^2)$ for $\Gamma\gg1$ 
is the Doppler factor and $d\tilde\Omega=d\tilde{\mu} d\tilde{\varphi}$ is the differential solid-angle subtended by 
the emitting region relative to the central source. It is clear from
equation~(\ref{eq:tobs}) that for a given observed time $t$, photons originating from different angles $\tilde\theta$, corresponding 
to angles $0\leq\theta\leq\theta_j$, and radii $r$ contribute to the measured flux. Therefore, the integral 
over $d\tilde\mu$ in equation~(\ref{eq:Fnu}) must take into account the radiation arriving from an equal arrival time surface \citep[e.g.][]{GPS99,Granot+08}, which relates $r$ and $\tilde\mu$ through equation~(\ref{eq:tobs}) for a given $t_z$ and 
which extends radially from $\xi_{\rm min}$ for $\tilde\mu=-1$ to $\xi_{\rm max}$ for $\tilde\mu=1$. For a given dynamical 
evolution of the shell these limiting radii can be obtained by finding the roots of the following equations
\begin{equation}\label{eq:rminmax}
    \frac{ct_z}{r_d} = \left\{\int_0^{\xi_{\rm min}}\frac{1+\beta(\xi')}{\beta(\xi')}d\xi',\quad
    \int_0^{\xi_{\rm max}}\frac{1-\beta(\xi')}{\beta(\xi')}d\xi'\right\}~.
\end{equation}
In the early coasting stage, when $\Gamma(\xi)\approx\Gamma_0\gg1$, the above two limits simplify into 
$\{\xi_{\rm min},~\xi_{\rm max}\} \approx \{ct_z/2r_d,~2\Gamma_0^2ct_z/r_d\}$.

In the thin-shell approximation, depending on the nature of the problem, the outer integral in equation~(\ref{eq:Fnu}) 
can either be performed over $\tilde\mu\in[-1,1]$ or $\xi\in[\xi_{\rm min},\xi_{\rm max}]$. In the latter case, integration over $\xi$ can be 
implemented with a simple calculation of the jacobian, such that $d\tilde\mu=\vert d\tilde\mu/d\xi\vert d\xi$, where
\begin{eqnarray}
    \tilde\mu &=& \frac{1}{\xi}\left[\int_0^\xi\frac{d\xi'}{\beta(\xi')}-\frac{ct_z}{r_d}\right] \\
    \frac{d\tilde\mu}{d\xi} &=& \frac{1}{\xi^2}\left[\frac{ct_z}{r_d}+\frac{\xi}{\beta(\xi)}-\int_0^\xi\frac{d\xi'}{\beta(\xi')}\right]
    \label{eq:dmudr}
\end{eqnarray}

In order to perform the integral over the azimuthal angle $\tilde\varphi$, without loss of generality, the LOS is considered 
to lie in the $\hat x$-$\hat z$ plane ($\varphi=0$). This yields $\hat{n} = \hat{\tilde{z}} = \sin\theta_{\rm obs}\hat x+\cos\theta_{\rm obs}\hat z$, 
and the unit vectors spanning the plane of the sky (normal to the LOS), 
$\hat{\tilde{x}} = \cos\theta_{\rm obs}\hat x - \sin\theta_{\rm obs}\hat z$ and $\hat{\tilde{y}}=\hat{y}$. Then expressing any 
radial unit vector $\hat r$ in both coordinate systems and projecting it onto the $\hat z$ axis yields the general relation 
\begin{equation}\label{eq:phi}
    \cos\left[\tilde{\varphi}(\tilde\mu,\mu,\mu_{\rm obs})\right] = \frac{\tilde\mu\mu_{\rm obs}-\mu}{\sqrt{(1-\tilde\mu^2)(1-\mu^2)}}\ .
\end{equation}
For a spherical flow, the properties of the emission don't depend on $(\tilde\mu,\tilde\varphi)$, and therefore the 
observer receives emission from  $\Delta\tilde\varphi=2\pi$. 

\begin{figure}
    \centering
    \includegraphics[width=0.48\textwidth]{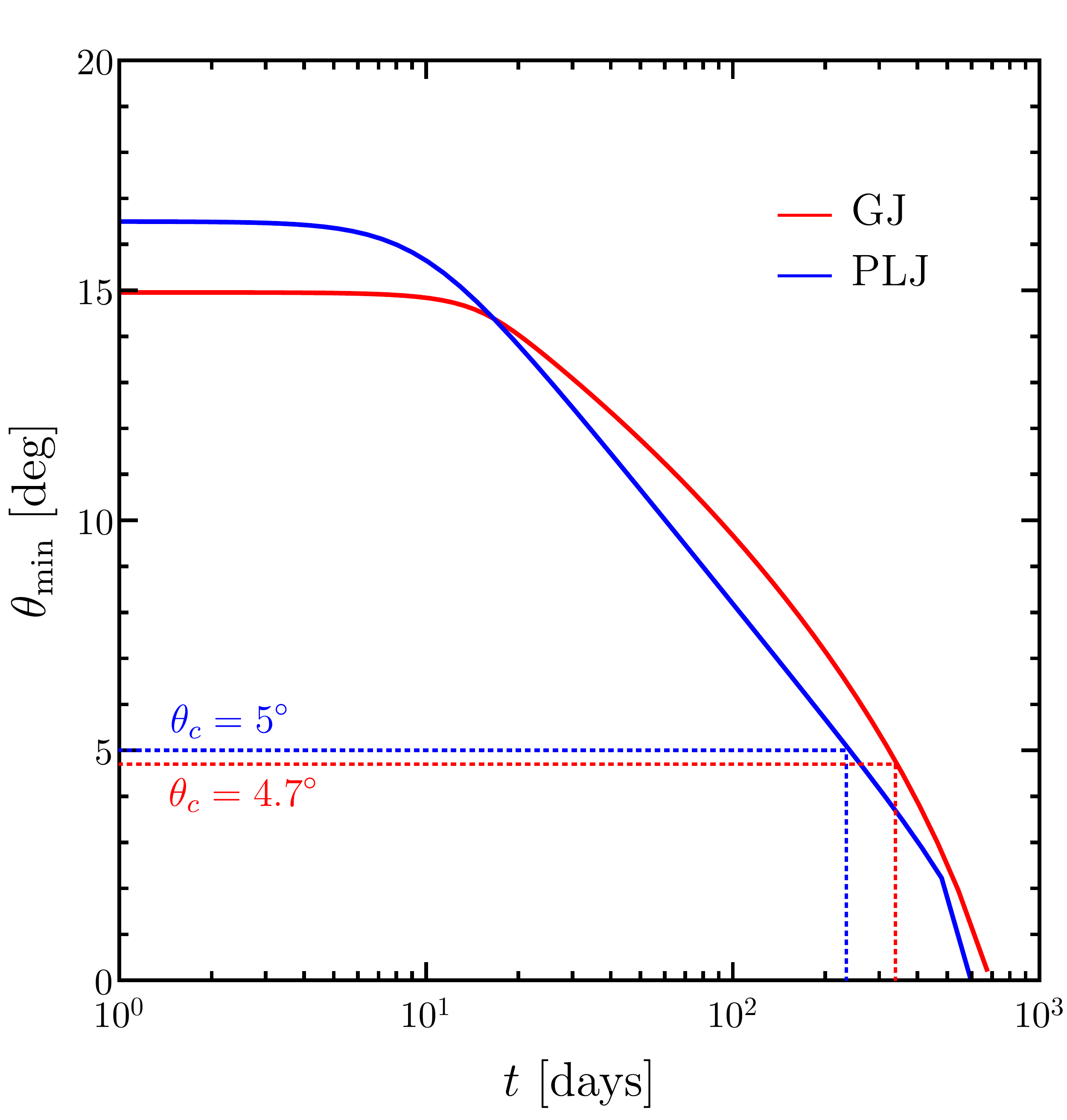}
    \caption{Minimum angle from which the emission contributes to the flux in the LOS.}
    \label{fig:theta-min}
\end{figure}

\subsection{Comparison of afterglow lightcurves with observations of GRB 170817A}

Here we compare the prediction of the lightcurves obtained for the structured jets 
\citep[e.g.][]{GK03,KG03,Rossi+04,LK17a,SGG18} and the (quasi-) spherical outflows \citep[e.g.][]{FV18,GNP18,Hotokezaka+18,SGG18}
to the radio (3~GHz and 6~GHz), optical (at 2~eV), and X-ray (at 1~keV) observations 
of SGRB~170817A \citep[e.g.][]{Margutti+18}. \textcolor{black}{The first X-ray and radio detections are at $8.9\;$days and $16.4\;$days, respectively, and the observed flux density at these wavelengths appears to be dominated by the afterglow emission throughout all of the observations so far. However, during the first few weeks the observed flux density in the optical (as well as in the IR and the early UV emission) is dominated by the kilonova emission. Therefore, in order to avoid any significant contribution of the kilonova component that is not included in our modeling, we use the optical observations only at sufficiently late times \citep{Lyman+18,Margutti+18} for fitting the simulated lightcurves.} Furthermore, all these observations, 
that were obtained between $\sim 9$ days and $\sim163$ days post merger, suggest that 
the afterglow radio to X-ray emissions lie on the same synchrotron power-law segment 
(PLS -- specifically PLS G as discussed in \citet{GS02}). This fact offers a way to 
constrain some of the parameters in the large parameter space of the models considered here. Therefore, 
all the lightcurves that are shown below respect the constraint that $\nu_m<3$~GHz and $h\nu_c>10$~keV over 
the entire period over which the afterglow data was obtained.

In the first row of Figure~\ref{fig:lc}, we show the afterglow lightcurves from the GJ and PLJ models. 
In both cases, the jet has a narrow core with $\theta_c\sim5^\circ$ and the viewing 
angle is $\theta_{\rm obs} \sim 27^\circ$. We stress that these are tentative fits 
to the data, which are by no means unique, and other sets of model parameters may 
provide a comparably good fit. Nonetheless, they are still representative for most 
purposes. In the second row of Figure~\ref{fig:lc}, we show the lightcurves for the 
SPh and QSph models. For these, we find that the values of $s$ are similar 
to the expected ones (compare to Appendix~\ref{sec:App-radial}). For the PLJ model, 
we obtain a value of $a=4.5$, and generally find that $a\gtrsim3.5-4$ is preferred 
by the afterglow data. This is significantly larger than the $a\approx2.7$ that is 
inferred from the asymptotic analytic estimate in Appendix~\ref{sec:App-angular}. 
However, this is likely due to the fact that in our case $\theta_{\rm obs}/\theta_c=5.4$ 
does not quite allow to reach the asymptotic range of $\theta$-values, 
$\theta_c<\theta\ll\theta_{\rm obs}$, for which that analytic estimate was calculated.

The (asymptotic) temporal index of $F_\nu(t)$,
$\alpha\equiv d\log F_\nu/d\log t$, is derived for a power-law $E_{\rm k,iso}(>u)\propto u^{-s}$ 
radial energy injection and for a jet with a narrow core and power-law wings in 
Appendixes~\ref{sec:App-radial} and \ref{sec:App-angular}, respectively. Figure~\ref{fig:fnu-slope} 
shows $\alpha(t)$ of the lightcurves from our tentative fits to the 
data. At early times, when the outflow is in the 
coasting phase, $F_\nu\propto R^3\propto t^3$. After the flow decelerates, marked by 
the decrease in the temporal index, the slight curvature in the lightcurves for 
$10~{\rm days}\lesssim t \lesssim 100$~days is apparent. The lightcurves in all models 
reach the peak at approximately the same time \textcolor{black}{at $t_p\sim150$~days \citep[also see][]{Dobie+18,Margutti+18}.} 
For $t>t_p$, the 
temporal index of the two structured jet models is steeper than that of any wide-angle 
quasi-spherical flow. This can potentially serve as a discriminator between the two 
types of jet profiles. For $t\gtrsim10^3$~days, the counter-jet starts to contribute to the 
flux and produces a flattening in the lightcurves.
Numerical simulations suggest that the counter-jet may have a stronger effect on the 
lightcurve when it becomes visible \citep{DeColle+12,GDCR-R18}.

\begin{figure*}
    \centering
    \includegraphics[width=0.45\textwidth]{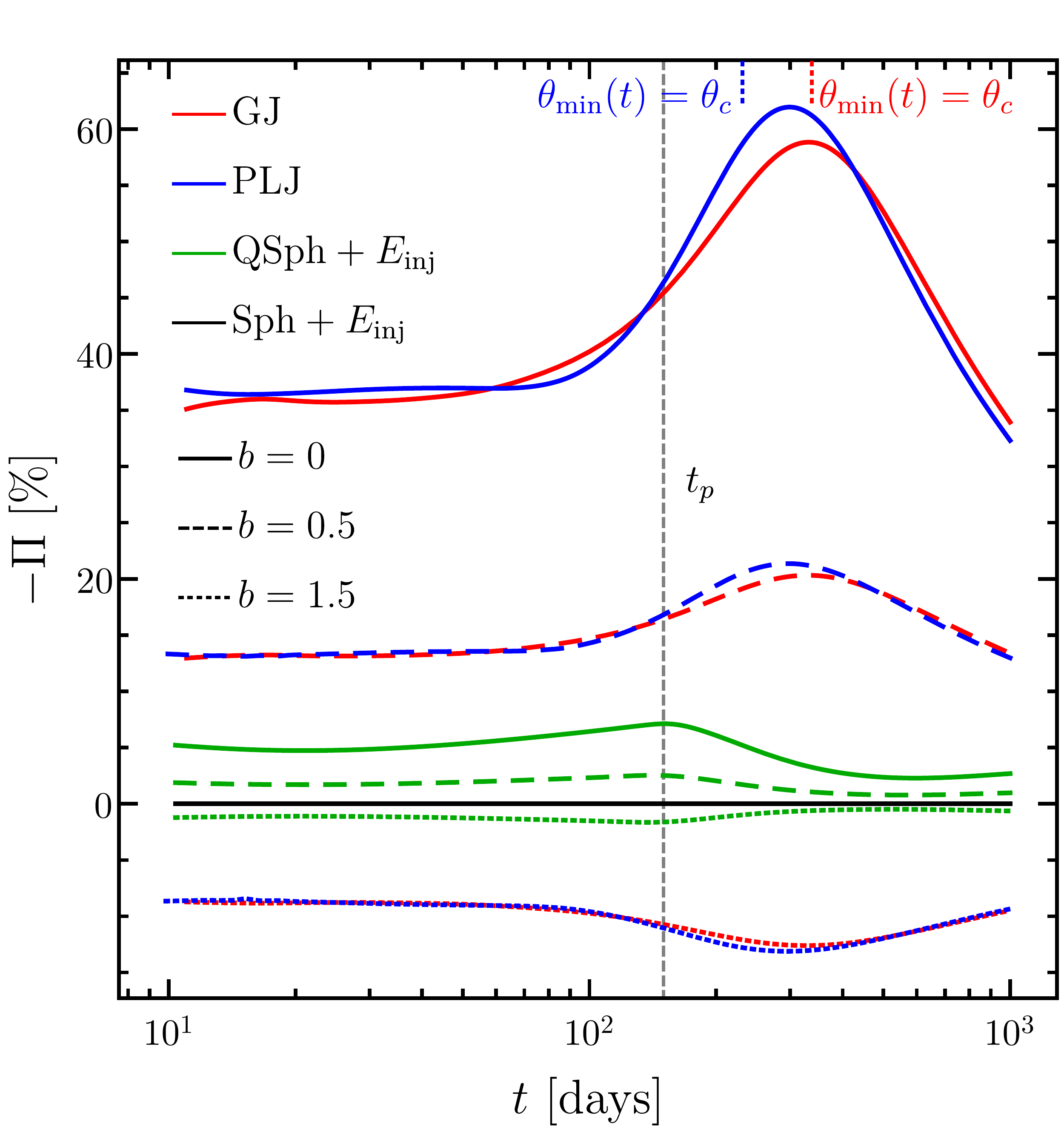}\hspace{1cm}
    \includegraphics[width=0.45\textwidth]{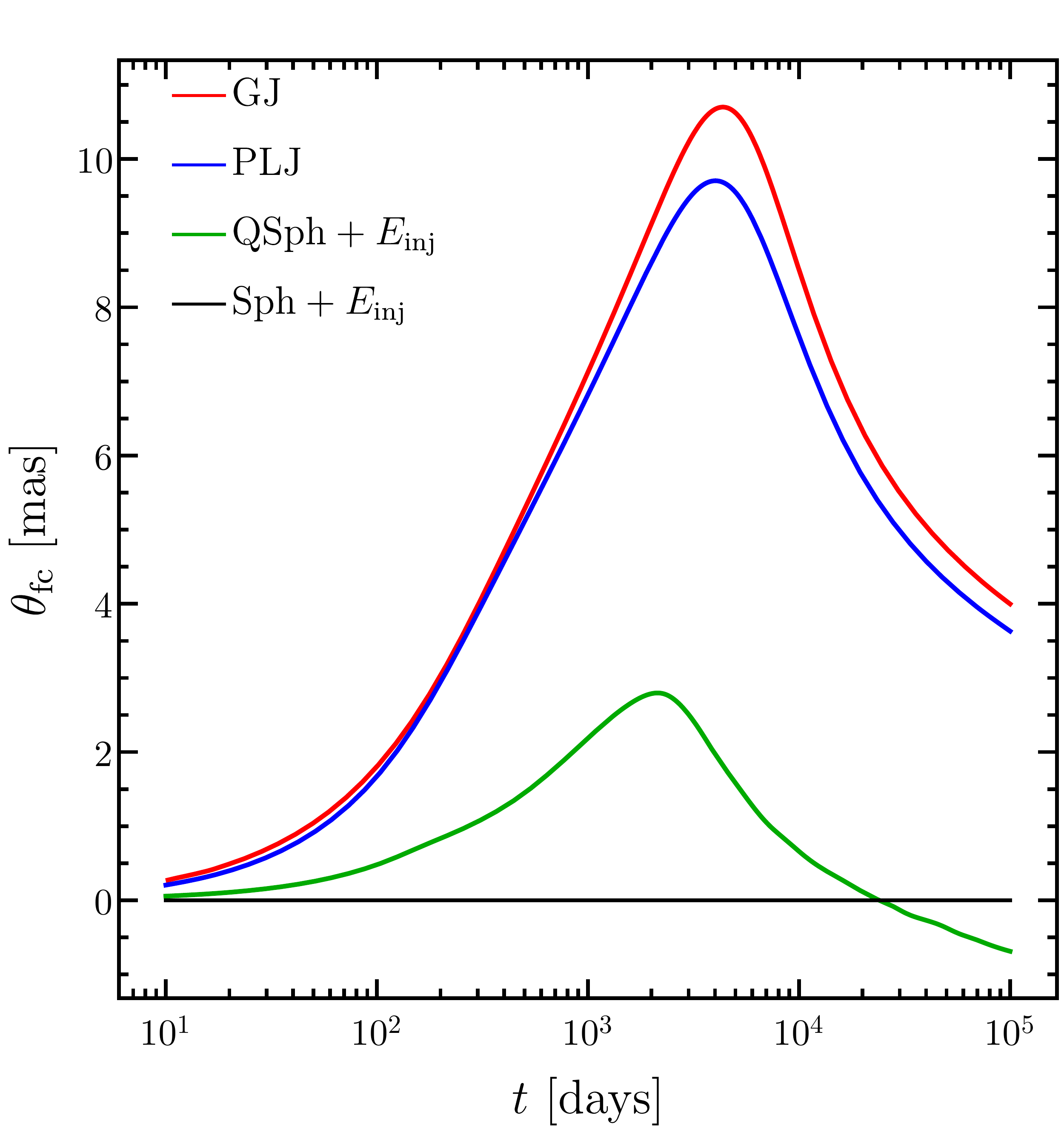}
    \caption{(Left) Linear polarization in radio at 3~GHz for the different models shown in this work. 
    The degree of polarization for different levels of magnetic field anisotropy is shown, where for
    $b=0$ (solid) the field is completely in the plane perpendicular to the shock normal, 
    and for $b=0.5$ (dashed) and $b=1.5$ (dotted) the field component in the direction of the shock normal 
    also contributes. For $b=1$ both components contribute equally, which yields no net polarization. Between 
    the two cases where $b<1$ and $b>1$, the polarization position angle changes direction by $90^\circ$. 
    The times at which emission from the jet core ($\theta=\theta_c$) starts contributing to the flux 
    are indicated with dotted lines. The peak time of the lightcurves $t_p$ is shown with a gray vertical 
    dashed line. (Right) Angular separation of the flux centroid (in milliarcsecond) from the center of the GRB.}
    \label{fig:lin-pol-fc}
\end{figure*}

In the case of structured jets, for which our fit parameters are very similar for the 
GJ and PLJ models, we compare our results to the the fit parameters 
obtained using MHD simulations in recent works. In \citet{Lazzati+17b}, the best fit parameters are: 
$\theta_{\rm obs}\sim33^\circ$, $\theta_c\sim1^\circ$, $E_{\rm iso,c}\sim10^{52}$~erg, 
$\Gamma_c\approx80$, $n_0\sim4\times10^{-3}~{\rm cm}^{-3}$, $\epsilon_e\approx6\times10^{-2}$, 
$\epsilon_B\approx3.3\times10^{-3}$, and $p \approx 2.07$. Likewise, in \citet{Margutti+18}, 
the best fit parameters (for one of their models) are: 
$\theta_{\rm obs}=20^\circ$, $\theta_c\sim9^\circ$, $E_{\rm iso,c}\sim10^{53}$~erg, 
$\Gamma_c\approx10^2$, $n_0=10^{-4}~{\rm cm}^{-3}$, $\epsilon_e=2\times10^{-2}$, 
$\epsilon_B=10^{-3}$, and $p = 2.16$. On the other hand, for the wide-angle 
flow scenario explored by \citet{Mooley+18}, one of their models, which is somewhat 
closer to the QSph model explored in this work, yields the following best fit parameters: 
$u_{0,\rm min} = 1$, $u_{0,\rm max} = 3.5$, $s=5$, $E_{\rm iso}=2\times10^{51}$~erg, 
$n_0 = 8\times10^{-5}~{\rm cm}^{-3}$, $\epsilon_e = 10^{-2}$, $\epsilon_B=10^{-1}$, 
and $p=2.2$. In a recent work, \citet{Resmi+18} conducted an MCMC maximum likelihood 
analysis using a semi-analytic model of a gaussian jet, much similar to the 
GJ model presented here, and obtained the following fit parameters: 
$E_{\rm k,iso,c}=10^{51.76}$~erg, $\Gamma_c=215$, $\theta_c=6.9^\circ$, 
$\theta_{\rm obs}=27^\circ$, $n_0=10^{-2.68}~{\rm cm}^{-3}$, $\epsilon_e=10^{-0.66}$, 
$\epsilon_B=10^{-4.37}$, and $p=2.17$. These model parameters are consistent with our results.

\section{Linear polarization}

\begin{figure*}
    \centering
    \includegraphics[width=0.48\textwidth]{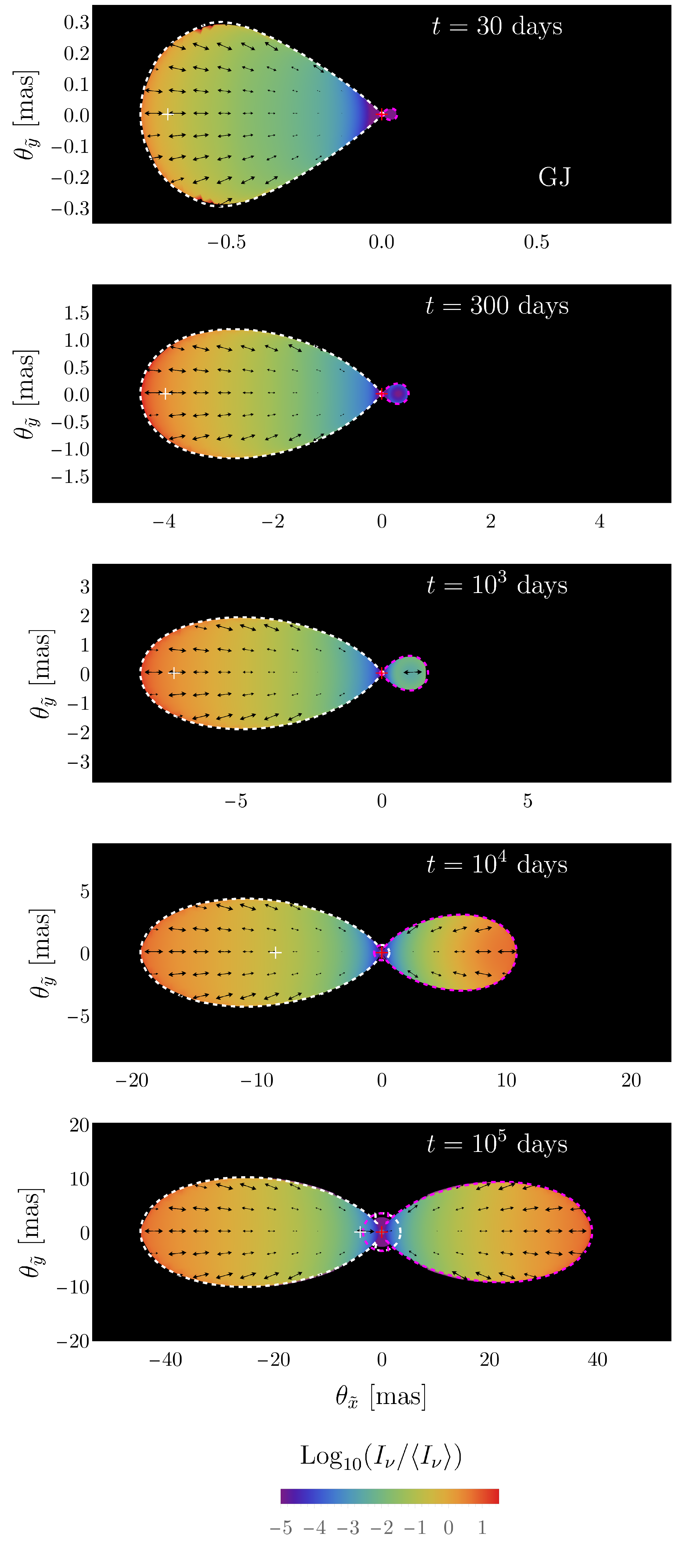}\quad\quad
    \includegraphics[width=0.48\textwidth]{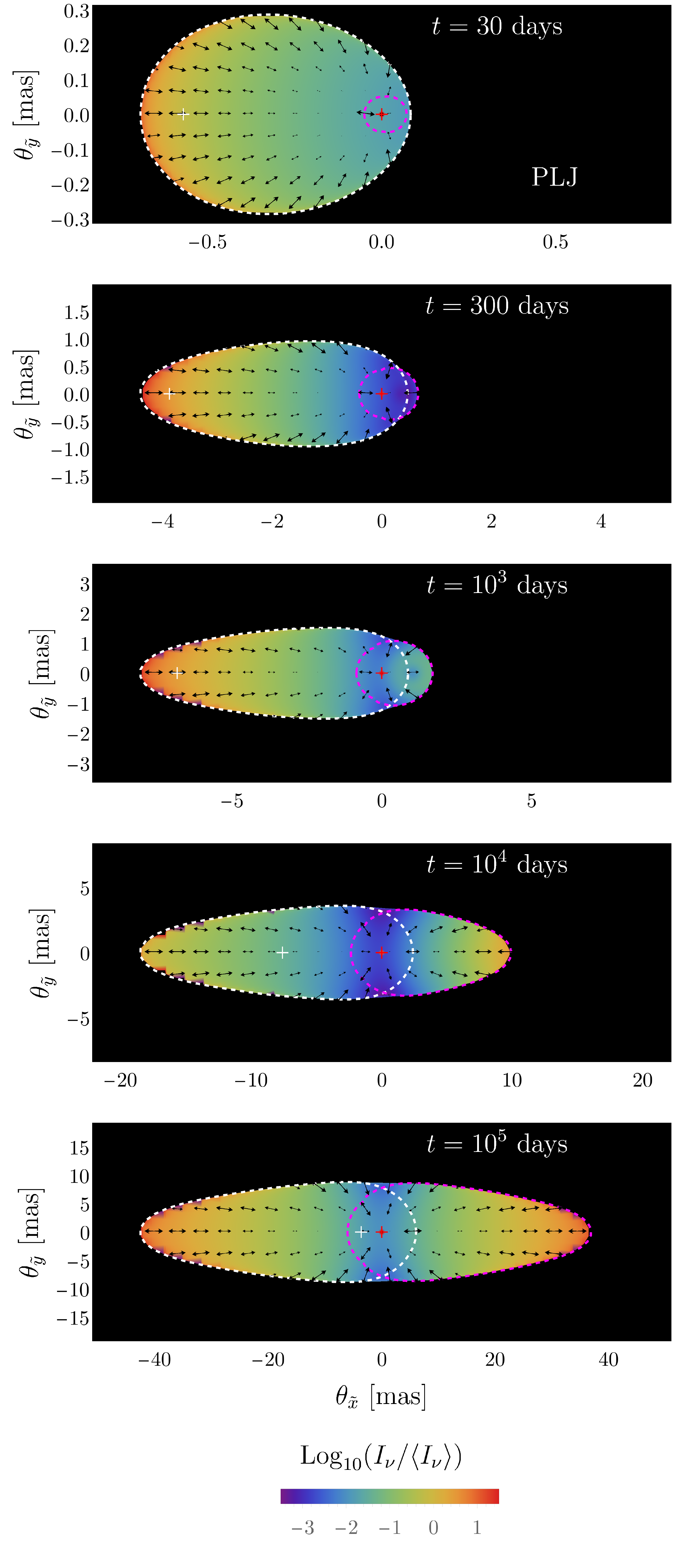}
    \caption{Radio afterglow images of the outflow on the plane of the 
    sky with polarization maps for the two different models discussed in this work: a Gaussian 
    jet (GJ; {\it left panels}) and a core+power-law jet 
    (PLJ; {\it right panels}). These normalized images are independent 
    of frequency within the same spectral PLS, and are shown here for
    PLS G where $F_\nu\propto\nu^{(1-p)/2}$. The maximum extents of the 
    main and counter jets are shown with white and magenta dotted lines, 
    respectively. The location of the GRB central source  is marked with 
    a red `+'-sign and the position of the flux centroid is marked 
    with a white `+'-sign. The polarization vectors are shown with double-sided 
    black arrows, whose length scales linearly with the degree of polarization.}
    \label{fig:SJ-images}
\end{figure*}

\begin{figure*}
    \centering
    \includegraphics[width=0.45\textwidth]{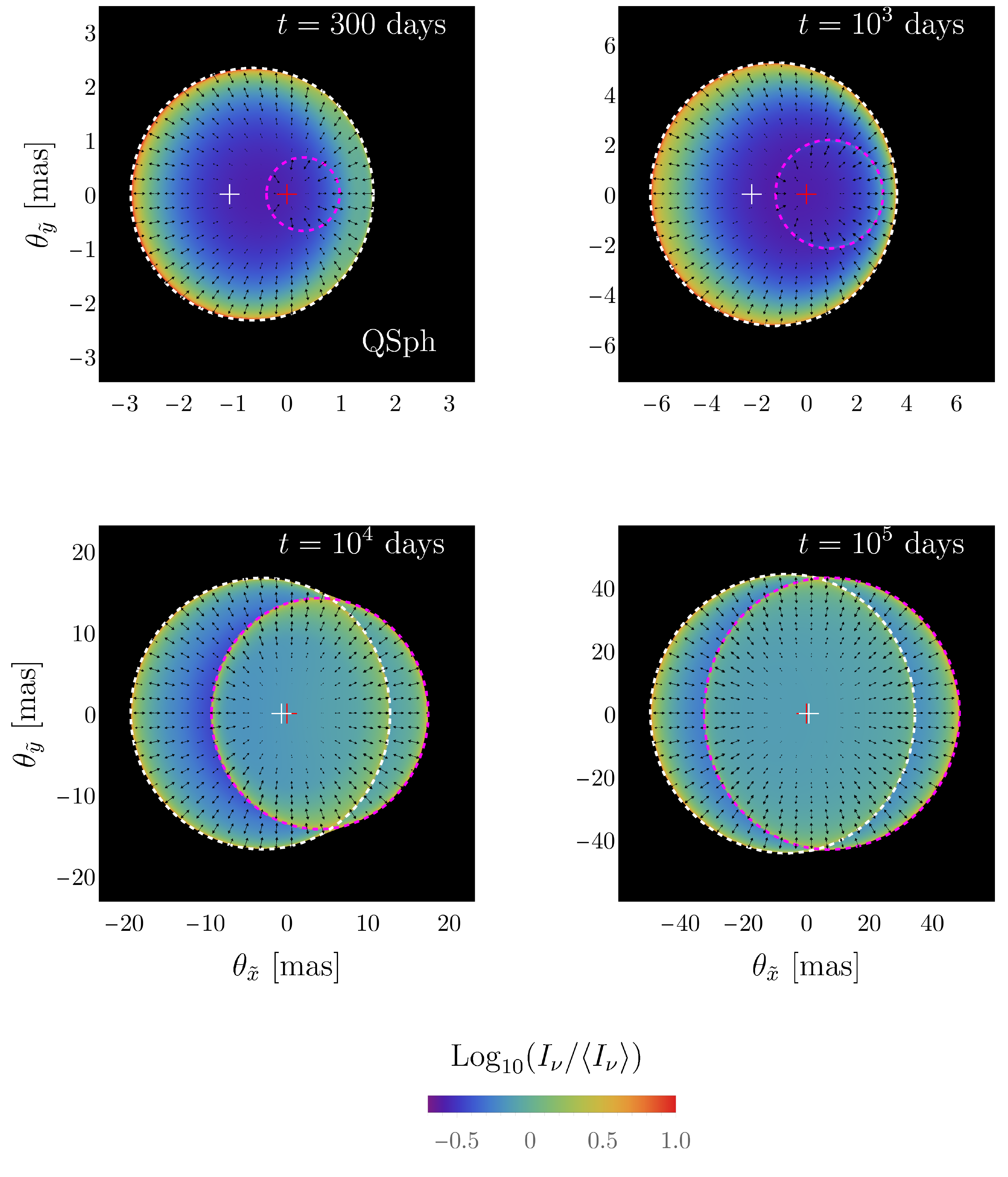}\quad\quad
    \includegraphics[width=0.45\textwidth]{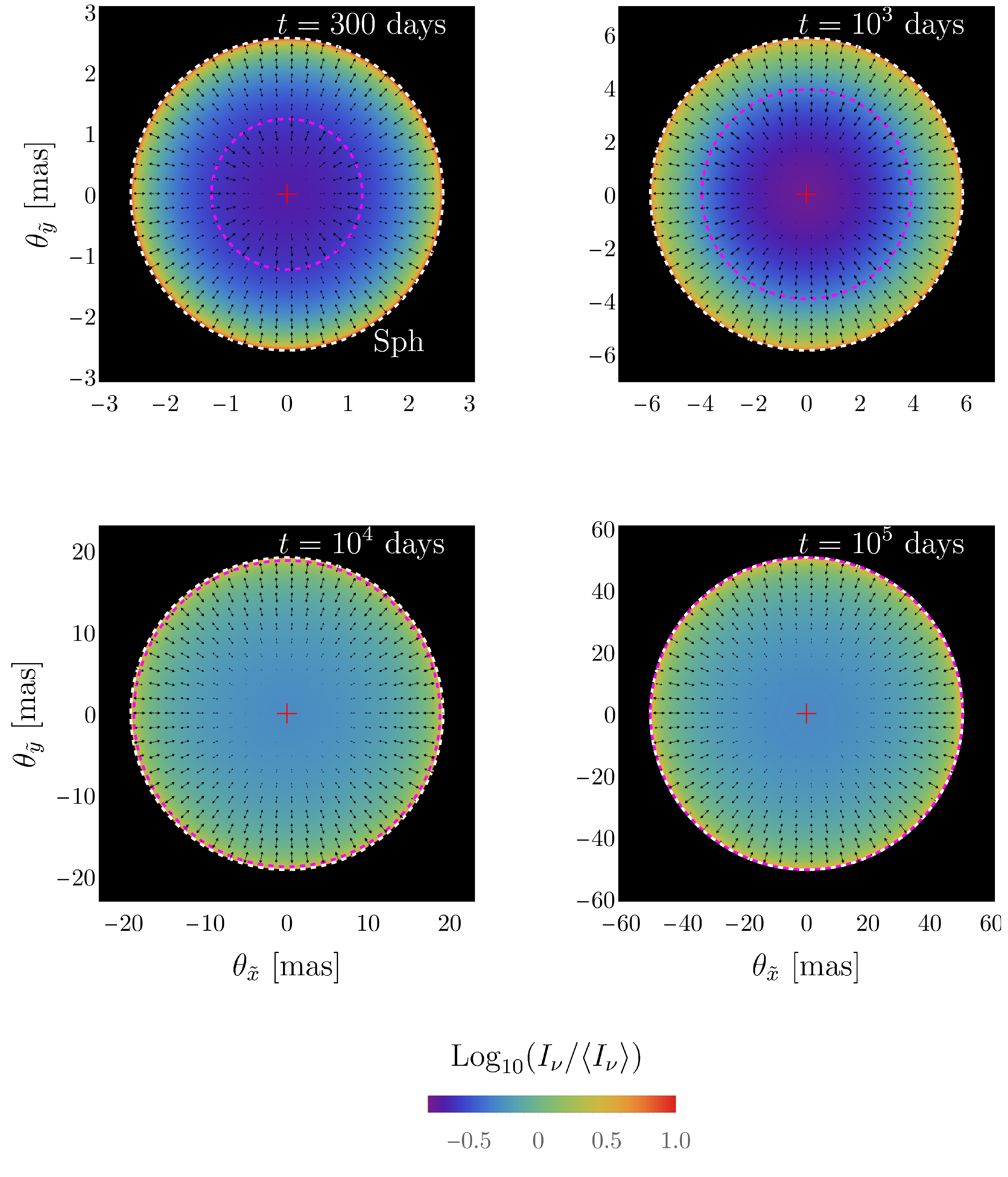}
    \caption{Radio afterglow images of the outflow on the plane of the sky with polarization maps 
    for the quasi-spherical shell (Qsph; left) and spherical shell (Sph; right) 
    with energy injection models. The format is the same as in Fig.~\ref{fig:SJ-images}.}
    \label{fig:Qsph-images}
\end{figure*}

The degree of linear polarization depends on the orientation of the magnetic field with respect to the 
LOS and its coherence length when compared with the angular size of the visible region, i.e. $\theta\sim1/\Gamma$. 
An ordered magnetic field with a large coherence length can give rise to a large degree of linear polarization $\Pi$ 
\citep{Granot03,Granot-Konigl03}. A more random magnetic field with a smaller 
coherence length would produce a smaller $\Pi$. A completely random field in 3D produces no net polarization even locally (over a region much larger than its coherence length but much smaller than the size of the emitting region). A completely random magnetic field in the plane of the shock produces local polarization in different parts of the image, but for a spherical flow it averages out to zero over the whole image (for an unresolved source such an effective averaging cannot be avoided) leaving no net linear polarization ($\Pi=0$). 
In this case, the axial symmetry around our line of sight needs to be broken. 
In our case this happens if the flow is axi-symmetric and our viewing angle 
is offset (by an angle $\theta_{\rm obs}>0$) relative to the flow's symmetry 
axis. This has been studied for a uniform jet with sharp edges that is viewed 
off-axis \citep[e.g.][]{Sari99,GL99}, or for an outflow with more angular 
structure, viz. $\epsilon(\theta)$ and/or $\Gamma(\theta)$ viewed off-axis 
\citep[e.g.][]{Rossi+04}.

Here we consider a random magnetic field that is tangled on angular scales $\ll1/\Gamma$, with 
axial symmetry with respect to the shock normal $\hat n_{\rm sh}$. The field anisotropy is parameterized 
by $b\equiv2\langle B_\parallel^2\rangle/\langle B_\perp^2\rangle$, where $B_\parallel$ ($B_\perp$) is 
the magnetic field component parallel (perpendicular) to $\hat n_{\rm sh}$ \citep{Granot-Konigl03}. 
In this case the local polarization in the comoving frame around the LOS is given by 
\citep{Gruzinov99,Sari99}
\begin{equation}\label{eq:Pi}
    \Pi(\tilde\theta') = \Pi_{\rm max}\frac{(b-1)\sin^2(\tilde\theta')}{2+(b-1)\sin^2(\tilde\theta')}
    \quad{\rm with}\quad\Pi_{\rm max} = \frac{p+1}{p+7/3}~,
\end{equation}
where $\Pi_{\rm max}\simeq0.7$ for $p=2.16$. For $b>1$ ($b<1$), the local polarization 
is $\Pi>0$ ($\Pi<0$) and the direction of the polarization vector is along (normal to) 
the direction of $\hat n\times\hat n_{\rm sh}$. The polar angles in the comoving frame can be 
related to that in the lab frame through the aberration of light,
\begin{equation}
    \tilde\mu' = \frac{\tilde\mu-\beta}{1-\beta\tilde\mu}~.
\end{equation}
Recall that the local dynamics depend on the polar angle from the jet symmetry axis, so that 
$\beta = \beta(\theta,r) = \beta(\tilde\mu,\tilde\phi,r)$. The degree of polarization can be conveniently 
expressed using the Stokes parameters, which are obtained by averaging over the polarization emerging 
from each fluid element in the visible region, such that
\begin{equation}
    \left\{
    \begin{array}{c}
        Q/I  \\
        U/I
    \end{array}
    \right\} = \frac{\displaystyle\int \tilde\delta_D^3L'_{\nu'}\Pi
    \left\{\begin{array}{c}
         \cos2\tilde\phi  \\
         \sin2\tilde\phi
    \end{array}\right\}d\tilde\Omega}
    {\displaystyle\int \tilde\delta_D^3L'_{\nu'}d\tilde\Omega}~.
\end{equation}
The degree of polarization is obtained from $\Pi = \sqrt{Q^2+U^2}/I$, where $I\propto F_\nu$. 

Let us first consider a structured jet, either GJ or PLJ. In this case, $\Pi$ rises 
with time as emission from more energetic regions at $\theta<\theta_{\rm obs}$ comes into view. The local beaming cone has a half-opening angle $\theta_b=\arccos{\beta}$ that approaches $1/\Gamma$ for $\Gamma\gg 1$. In 
Figure~\ref{fig:theta-min}, we show the minimum polar angle $\theta_{\rm min}$ from the jet symmetry axis 
the beaming cone of which just includes the LOS, i.e. when $\theta_{\rm obs}-\theta_{\rm min}=\theta_b(\theta_{\rm min})=\arccos{\beta}\approx1/\Gamma(\theta_{\rm min})$, which in the relativistic regime corresponds to $\Gamma(\theta_{\rm min})(\theta_{\rm obs}-\theta_{\rm min})\approx1$. 
Initially, at early times during the coasting phase ($t\lesssim 10\;$days in our case) $\theta_{\rm min}$ assumes a constant value.
Over time, as the faster moving parts of the jet slow down and their beaming cones widen, $\theta_{\rm min}$ gradually decreases and moves closer to 
$\theta_c$. Since most of the energy resides at $\theta\lesssim\theta_c$, the level of polarization peaks, 
as shown in the left panel of Figure~\ref{fig:lin-pol-fc}, near the time when $\theta_c$ becomes visible to the 
observer (around $\sim300\;$days or so in our case).
This is a factor of $\sim2$ after the peak of the lightcurve, $t_p$, since a rise in flux in such an angular scenario requires a sufficiently fast increase in the energy within the visible region (see Appendix~\ref{sec:App-angular}), but $\epsilon(\theta)$ starts to level off already somewhat above $\theta_c$ so that the flux peaks and starts to decay when $\theta_{\rm min}$ is still somewhat larger than $\theta_c$. Such an effect is not seen for the quasi-spherical model, where the time of the peak in the lightcurve and in the polarization practically coincide.

The degree of polarization starts to decline as the observed flux is dominated by the jet's core, which continues to decelerate so that the photons that reach us are emitted closer to the shock normal in the comoving frame (at smaller $\theta'$), which reduces $\Pi(\theta')$  (see Eq.~(\ref{eq:Pi})). The polarization and its time evolution, $\Pi(t)$, is very similar for our two off-axis structured jet models.

The linear polarization, in particular near the time of the peak in the lightcurve, 
is much larger for an off-axis jet whose energy is dominated by its narrow core, 
compared to a quasi-spherical flow. However, the degree of polarization for all 
outflow structures considered here could decrease by about the same factor if in 
reality the magnetic field behind the afterglow shock is not random only fully 
within the plane of the shock ($b=0$) \citep[e.g.][]{Sari99,Granot03,Granot-Konigl03}, 
but also has a comparable random component in the direction of the shock normal ($b>0$).
This might be hinted by the relatively low levels of linear polarization usually 
measured in GRB afterglows in the optical or NIR \citep[of $\Pi\lesssim\;$a few \%, e.g.][]{CG16}.
Therefore, a potentially more robust difference between the expected $\Pi(t)$ 
for these different outflow models is its time evolution -- for our off-axis 
jets there is a more distinct peak in $\Pi(t)$ near the time of the peak in the 
lightcurve, $t_p$.

The polarization vector on the plane of the sky is expected to be along the $\tilde{x}$-axis 
(which is also along the direction of motion of the flux centroid) for $0\leq b<1$, and along 
the $\tilde y$-axis (which is normal to the direction of motion of the flux centroid) for 
$b>1$.

\section{The radio image -- flux centroid, size \& shape}

Possibly the most promising way to break the degeneracy between the models considered 
in this work is by comparing the properties of the image on the plane of the sky, 
especially in radio \citep[see e.g.][for a review]{GV14}, to that obtained from the 
various models. Several properties of the radio image can potentially be directly 
compared with observations, depending on whether and how well the image is resolved. 
Another important diagnostic that can help break the degeneracy between different 
outflow models is the motion of the flux centroid \citep[e.g.][]{Sari99,GL03}, which 
may in some cases be detected even if the image is only marginally resolved or even not 
resolved altogether. 

In order to calculate the flux centroid, we consider the image of the outflow on the 
plane of the sky with coordinates $(\tilde x,\tilde y)$, where the 
line connecting the LOS to the jet symmetry axis coincides with the $\tilde x$-axis. The image will always be 
symmetric around this line, and therefore the flux centroid will move along the $\tilde x$-axis. The position 
of the flux centroid $\tilde x_{\rm fc}(t)$, expressed in terms of the angular displacement 
$\theta_{\rm fc}(t)$ from the location of the GRB central source, is simply an average of 
$\tilde x = \tilde\rho\cos\tilde\phi=R\sqrt{1-\tilde\mu^2}\cos\tilde\phi$ weighted by $F_{\nu}$, such that
\begin{equation}
    \tilde x_{\rm fc}(t) = \frac{\int \delta_D^3L'_{\nu'}\tilde{\rho}\cos\tilde{\phi}\, d\tilde\Omega}
    {\int\delta_D^3L'_{\nu'}\,d\tilde\Omega}\quad{\rm and}\quad
    \theta_{\rm fc}(t)\equiv\frac{\tilde x_{\rm fc}}{d_A}\approx\frac{\tilde x_{\rm fc}}{d}~,
\end{equation}
where $d_A = (1+z)^{-1}d$ is the angular distance and $d$ is the proper distance, and $d_A\approx d$ when $z\ll1$; we use
$d=40$~Mpc for GRB~170817A.

In the right panel of Figure~\ref{fig:lin-pol-fc}, we show the motion of the 
flux centroid for all the models considered in this work. The angular position 
of the flux centroid evolves in a similar way for the GJ and PLJ models. 
However, its evolution is very different for the QSph model (for the Sph model 
it obviously does not move at all). The maximum $\theta_{\rm fc}(t)$ for the 
QSph model is significantly lower than that predicted for the two structured 
jets, and it peaks (i.e. its movement reverses direction) at a slightly earlier 
time compared with GJ and PLJ models. For the GJ and PLJ models $\theta_{\rm fc}(t)$ 
peaks around the time when the jet's core becomes sub-relativistic and the 
counter-jet's core becomes visible.


In Figures~\ref{fig:SJ-images}~\&~\ref{fig:Qsph-images}, we show the radio images on the 
plane of the sky for the GJ, PLJ, QSph, and Sph models \citep[also see][]{Nakar+2018}.
The specific intensity $I_\nu$ is normalized by its mean value within the image, $\mean{I_\nu}$, and these normalized images are independent of frequency within the same spectral PLS, and are shown here for PLS G where $I_\nu,\,F_\nu\propto\nu^{(1-p)/2}$. Since the emission is from a thin shell the images are particularly limb brightened and $I_\nu$ diverges near the outer edge of the image as the square root of the projected distance from the edge \citep{Sari98,GL01,Granot08}. When the emission from the bulk of the hot plasma behind the afterglow shock is considered the resulting images are somewhat less limb brightened, and the surface brightness no longer diverges and instead peaks at a lower value somewhat before the outer edge of the image \citep{GPS99,GL01,Granot08}. However, for PLS G with $p=2.2$, $I_\nu$ within the circular image for the \citet{BM76} self-similar solution peaks at 95\% of its outer radius, and the overall limb brightening is not that different from PLS H for which the emission is indeed from a very thin cooling layer just behind the shock \citep[see Fig.~2 of][]{Granot08}.

At early times, the emission from the main jet 
dominates the intensity and observed flux, and therefore determines the location of the flux centroid, while emission from the 
counter-jet is beamed away from the observer. At late times, when the counter-jet's core becomes 
sub-relativistic the counter-jet's contribution to the observed flux becomes more prominent and it 
starts moving the flux centroid back towards the location of the central source.

\begin{figure}
    \centering
    \includegraphics[width=0.48\textwidth]{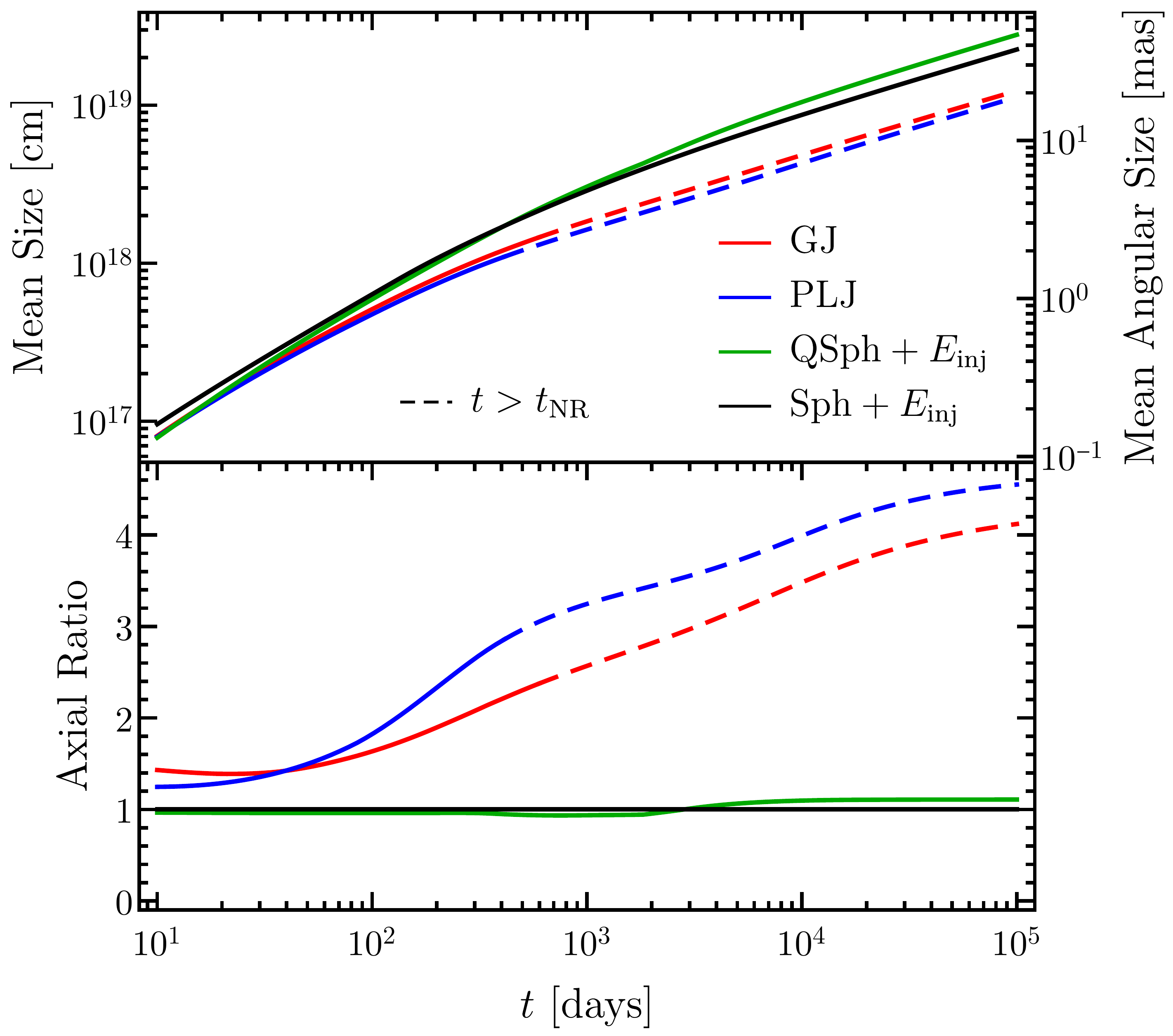}
    \caption{(Top) Geometric mean size of the image (half of the total image 
    extent in the $\tilde x$ and $\tilde y$ directions). (Bottom) Axial ratio 
    of the image. The dashed line shows the mean size and axial ratio at times 
    after the main jet becomes non-relativistic ($t>t_{\rm NR}$) for the 
    GJ and PLJ models. Due to the simple lateral dynamics (no lateral spreading) 
    assumed in this work, the predictions for the size and axial ratio may not 
    be so robust.}
    \label{fig:size}
\end{figure}

We show the evolution of the mean size of the radio images and its axial ratio over time in Figure~\ref{fig:size}. The difference in the angular size near the peak  of the lightcurve $t_p$ between the different models is rather modest ($\lesssim25\%$), and even at very late times it is a factor of $\lesssim2$. Therefore, this may not be the best way to distinguish between the different models. However the image axis ratio, which parameterizes its degree of elongation may be a better and more robust way to distinguish between the two main types of models (GJ or PLJ versus QSph or Sph). For the (quasi-) spherical model the image is (almost) circular, while for the off-axis structured jet models the image is rather elongated with an axis ratio of $\gtrsim2$ near $t_p$, and even somewhat more elongated at later times.

Since GRBs are usually cosmological sources and their afterglow images may at best be only marginally resolved, it is challenging to measure the actual angular size or shape of the outflow from radio observations. Instead the visibility data is fit to an assumed parameterized image surface brightness distribution.
The surface brightness of these sources is often modeled as a circular or an elliptical Gaussian \citep[e.g.][]{Taylor+05,TG06,Pihlstrom+07,Mesler+12}.
The results of such a fit may be biased due to the inhomogeneous brightness profile of the outflow, if it is significantly different than the assumed functional form. Therefore, the ouflow 
sizes inferred from e.g. radio images of GRBs may potentially be somewhat smaller than their true full sizes \citep[e.g.][]{Taylor+04,Pihlstrom+07}.

Our simplified dynamics may introduce some differences in the resulting afterglow images 
compared to more realistic full hydrodynamic simulations. Our neglect of the lateral jet 
dynamics affects mainly our jet models (our spherical model does not suffer from this problem, and the expected effects for the quasi-spherical model are also rather modest). It may render the results less realistic at late times, especially when the flow 
becomes Newtonian and is expected to approach the spherical self-similar Sedov-Taylor 
solution. Therefore, the relatively large image axis ratio at very late times, well after 
the jet's core becomes sub-relativistic at $t_{\rm NR}\sim2\,-\,3\;$yr,
is likely to be more modest and gradually decrease with time rather than increase with time as 
the flow becomes more spherical and Newtonian at such late times. It is more reasonable to 
expect the image axis ratio to peak at $\sim 2\,$--$\,3$, around $t_{\rm NR}$, and then gradually decrease. 

An accompanying paper \citep{GDCR-R18} presents afterglow images from 2D relativistic hydrodynamic 
simulations of an initially conical jet with sharp edges. One can get a better idea of the expected 
differences between our simplified dynamics and hydrodynamic simulations by a comparison with those 
results, despite the different initial jet structure.
For our simplified dynamics the jet's non-relativistic transition radius $R_{\rm NR}$ and corresponding 
observer time $t_{\rm NR}\sim R_{\rm NR}/c$ is given by the Sedov radius corresponding to $E_{\rm k,iso}(\theta=0)$. 
Alternatively one can estimate it for semi-analytic models featuring exponential lateral expansion at $R>R_j$ 
where the jet break radius $R_j$ is approximately the Sedov radius corresponding to the jet's true energy, 
which gives $R_{\rm NR}\approx(1-\ln\theta_0)R_j$. The ratio of the latter and former radii is 
$\theta_0^{2/(3-k)}(1-\ln\theta_0)$ which for $\theta_0=0.2$ gives 0.89 for a uniform density ($k=0$) 
but $0.10$ for a wind-like external density profile ($k=2$). While simulations give a result closer to 
the latter radius for a wind-like density profile, for the purposes of this work a uniform density is 
relevant, for which there are very small differences between these two estimates, so that our results 
should be quite reasonable. For a uniform density this time and radius scale as $t_{\rm NR}\sim R_{\rm NR}/c\propto(E/n)^{1/3}$
with the jet energy and external density.
The angular scale of the image around this time is $\sim\theta_{\rm NR}=R_{\rm NR}/d$.

For the hydrodynamics presented in \citet{GDCR-R18} the counter jet dominates the observed flux just after it becomes visible, causing the flux centroid to move to the other side of the central source, reaching a maximum displacement on that (counter-jet's) side a factor of $\sim 2$ in time after it passes through the projected location of the central source ($\tilde{x}=0$), and then gradually moves back towards it at late times as the contribution of the main and counter jets becomes closer to each other. In this work the effect of the counter jet is smaller, and it never quite dominates the flux. This likely occurs due to the more gradual deceleration of the jet's core in our simplified dynamics. A similar trend appears for a wind-like external density profile in the simulations \citep{DeColle+12}, for which the jet decelerates more slowly.
 

\section{Discussion \& Conclusions}
The broadband afterglow lightcurve of GRB 170817A that continues to rise even $\gtrsim 115$~days post merger has 
seriously challenged the naive view that outflows in GRBs are narrowly beamed and have sharp edges, with perhaps a homogeneous angular profile. Such a picture is also inconsistent with the sub-luminous prompt gamma-ray emission of GRB~170817A \citep[e.g.][]{GGG17}, which appears to arise from material along our line of sight. Model fits to radio and X-ray observations have revealed that the rising flux of GRB 170817A can be explained by two completely distinct models, namely a structured jet and a quasi-spherical outflow with initial radially stratified velocity profile. In terms of the lightcurves, both types of models can explain the current observations. The
predicted flux decay after the peak in the lightcurve is somewhat steeper for the off-axis structured jet models compared to the (quasi-) spherical models (see Figs.~\ref{fig:lc} and \ref{fig:fnu-slope}), but this may still not suffice to clearly distinguish between these models. Therefore, new types of diagnostics are 
needed to break this degeneracy.

In this work, we present three different diagnostics that appear to be most promising
and may also be observationally feasible, which may help to unveil the true nature of the outflow that powered GRB~170817A:
\begin{enumerate}
    \item \textit{Polarization}: The degree of polarization ($\Pi$) for the structured jets, namely the GJ and PLJ models, 
    undergoes a sharp increase beyond $\sim100$~days and peaks at $\sim300$~days, where $\Pi\approx60\%$ (for $b=0$). This trend 
    is in stark contrast with any wide-angle quasi-spherical flow for which $\Pi\lesssim10\%$. Radio or optical measurements of the 
    afterglow polarization may help distinguish between the structured jet and the `cocoon' scenario 
    \textcolor{black}{(also see e.g. \citet{D'Avanzo+18,Nakar+2018})}. A caveat here is that high $\Pi$-values assume a magnetic field that is 
    fully random within the shock plane ($b=0$), in 2D, while a field that is partly random in 3D and also has a comparable 
    component in the direction of the shock normal ($b>0$) could potentially significantly reduce $\Pi$, by a similar factor for 
    these different models. Another potential diagnostic may be obtained by comparing the peak time $t_\pi$ for $\Pi(t)$ and 
    $t_p$ for $F_\nu(t)$: for the GJ and PLJ models $t_\pi/t_p\approx 2$ while for a wide-angle quasi-spherical flow $t_\pi/t_p\approx 1$.
    \item \textit{Flux centroid motion}: A potentially powerful diagnostic is the motion of the flux centroid in relation to the 
    location of the GRB (that corresponds to the flux centroid's location at very early or late times). Both the GJ and PLJ models 
    show a large displacement of the flux centroid (reaching $\sim3$~mas at 
    $\sim200$~days) due to the modest viewing angle and the inherent angular profile of the outflow. On the other hand, a 
    lower offset ($\lesssim1$~mas at $\sim200$~days) is expected from any quasi-spherical flow.
    \item \textit{Axial ratio of the image}: The size of the image and its axial ratio, which may be determined using VLBI, can be 
    instrumental in discerning the properties of the outflow \citep[see e.g.][]{Taylor+05,TG06,Pihlstrom+07,Mesler+12}. We find that 
    all the models that are considered in this work and fit $F_\nu(t)$ the predicted image sizes as a function of time are approximately 
    similar. This makes it challenging to differentiate between the different models. However, the axial ratio can serve as an important 
    discriminator between a structured jet and quasi-spherical outflow at the current epoch. On the 
    other hand, the difference in the axial ratio between the GJ and PLJ models is $\lesssim 25\%$, 
    which remains approximately at that level even at late times. This makes it harder to distinguish 
    between the two jet profiles.
    The axial ratio for any wide-angle flow remains very close to unity at all times, while for the structured jets the axial ratio is $\sim2$ at $\sim200$~days.
\end{enumerate}

A high angular resolution instrument is needed to resolve the image and measure the flux centroid's movement, even for GRB~170817A that is at a relatively nearby distance of $\approx40\;$Mpc. The relatively small distance required for detecting binary mergers in gravitational waves is also more favorable for imaging compared to cosmological GRBs. In order to break the degeneracy between our structured jets and quasi-
spherical models for GRB~170817A at $\sim 200$~days, a minimum angular resolution of $\approx2$~mas is needed. This is within reach of the VLBA network in the northern hemisphere, where using the longest baseline there of 8008~km between the Effelsberg and the Jansky
Very Large Array (JVLA) site in New Mexico, the minimum angular resolution is $\sim170\;\mu$as at 43~GHz or $\approx1.5$~mas at 5~GHz. Since observations of GRB~170817A show that the flux density falls of with frequency as $F_\nu\propto\nu^{-0.6}$, it may not be possible to realize in practice the higher theoretical angular resolution at higher frequencies,
and instead it might be required to opt for lower frequencies despite the lower corresponding possible angular resolution, because of the higher $F_\nu$
that may hopefully enable to actually image and resolve the source in practice. The motion of the flux centroid may potentially be determined to somewhat better accuracy, and may possibly be measured even if the image is not resolved.

Linear polarization of GRB afterglow emission mostly at the $\sim1\% - 2\%$ level has been obtained for several GRBs in the optical or NIR, 
but only upper limits were obtained in radio \citep[e.g.][]{CG16}. The detection of polarization depends on the sensitivity of the 
instrument and the flux of the source, which together yields a measure of signal-to-noise (SNR). A high SNR is typically needed to 
register any polarization. Therefore, it may be challenging to measure the afterglow polarization for GRB~170817A.

In this work, we provide clear predictions for the afterglow lightcurves, polarization, and image  properties for the four different outflow models 
that can explain the observed flux evolution from radio to X-rays. Broadband observations in the near future may be able to distinguish between structured jet and quasi-spherical outflow models for GRB~170817A. 
This can take us one step closer to unraveling the nature of the outflows in GRBs.

    

\section*{Acknowledgements}
We thank the anonymous referee for useful suggestions. 
RG and JG acknowledge support from the Israeli Science Foundation under
Grant No. 719/14. RG is supported by an Open University of Israel Research Fund.






\onecolumn
\appendix
\subsection{Analytic scalings for a radial structure}
\label{sec:App-radial}
Here we consider a spherical outflow with a distribution of energy as a function of the 
ejecta's proper velocity, $E(>u)=E_0 u^{-s}$. For an external density $\rho = AR^{-k}$, 
the swept up mass within radius $R$ is $M(<R)=[4\pi/(3-k)]AR^{3-k}$. Beyond the deceleration 
radius ($R>R_d$), assuming that the flow expands adiabatically, we have $E(>u)=E_0 u^{-s}=M(<R)c^2u^2$
so that
\begin{equation}
u(R) = \left[\frac{(3-k)E_0}{4\pi Ac^2R^{3-k}}\right]^{1/(2+s)}\propto R^{-(3-k)/(2+s)}\ .
\end{equation}
In the relativistic regime $u\approx\Gamma\gg 1$ and the observer time (for $z\approx0$) 
$t\approx R/2c\Gamma^2$ so that
\begin{equation}
    \Gamma(t)\approx
    \left[\frac{(3-k)E_0}{2^{5-k}\pi Ac^{5-k}t^{3-k}}\right]^{1/(8-2k+s)}\propto t^{-(3-k)/(8-2k+s)}\ ,
\end{equation}
and $E_{\rm k,iso}\propto\Gamma^{-s}\propto t^{(3-k)s/(8-2k+s)}$. Since $R\propto\Gamma^2t\propto t^{(2+s)/(8-2k+s)}$ 
this implies that $F_\nu^{(G)}\propto \rho^{1/2}E_{\rm k,iso}^{(3+p)/4}t^{(3-3p)/4}\propto t^\alpha$ with 
$\alpha=[(3-k)s(3+p)-2k(2+s)]/[4(8-2k+s)]+(3-3p)/4$. An observed value of $\alpha$ could be reproduced by 
\begin{equation}
    s = \frac{2k+(8-2k)(4\alpha+3p-3)}{(3-k)(3+p)-2k-4\alpha-3p+3}\;\xrightarrow{k\rightarrow0}\;
    \frac{8\alpha+6p-6}{3-\alpha}\ ,
\end{equation}
which for the parameters relevant for GRB~170817A  ($k=0$, $p=2.2$, $\alpha=0.8$) implies $s\approx 6.2$. 

In the Newtonian regime $u\approx\beta\ll1$ and $R\sim\beta ct$ (for $z\approx0$) so that $\beta\propto t^{-(3-k)/(5-k+s)}$ 
and $R\propto t^{(2+s)/(5-k+s)}$. The peak flux scales as $F_{\nu,{\rm max}}\propto BN_e\propto\rho^{1/2}\beta  R^{3-k}\propto R^{3-3k/2}\beta$ while the typical synchrotron frequency scales as $\nu_m\propto B\gamma_m^2\propto\rho^{1/2}\beta^5\propto R^{-k/2}\beta^5$ leading to $F_\nu^{(G)}\propto F_{\nu,{\rm max}}\nu_m^{(p-1)/2}\propto t^\alpha$ with $\alpha = [(2+s)(12-5k-kp)-(6-2k)(5p-3)]/[4(5-k+s)]$ for which an observed value of $\alpha$ could be reproduced by 
\citep[also see e.g.][]{NP18}
\begin{equation}
    s = \frac{2(15p-21+10\alpha)-k[8(p-2)+4\alpha]}{4(3-\alpha)-k(5+p)}\;\xrightarrow{k\rightarrow0}\;
    \frac{15p-21+10\alpha}{6-2\alpha}\ ,
\end{equation}
which for the parameters relevant for GRB~170817A  ($k=0$, $p=2.2$, $\alpha=0.8$) implies $s\approx 4.5$. 

\subsection{Analytic scalings for an angular structure}
\label{sec:App-angular}
Here we consider a relatively simple angular structure of a jet with a narrow uniform core at an angle $\theta_c\ll 1$ with power law wings in $E_{\rm k,iso}=4\pi(dE/d\Omega)$ where $E_{\rm k,iso}(\theta>\theta_c)\propto\theta^{-a}$, viewed at an angle $\theta_{\rm obs}$ from its symmetry axis. For simplicity we assume that the jet retains its initial angular structure (i.e. we neglect lateral spreading) and assume that at each angle $\theta$ from the jet's symmetry axis the  flow evolves as if it were a part of a spherical flow with the local $E_{\rm k,iso}(\theta)$. In this scenario the jet is relativistic and gradually decelerates as it sweeps up the external medium.
At each observed time $t$ the parts of the jet that can significantly contribute to the observed emission are those whose beaming cone (of half-opening angle $1/\Gamma(\theta,t)$ around their direction of motion, which is assumed to be radial here) includes our line of sight. 
Therefore, they can be treated as viewed "on axis", and satisfy the usual on-axis relations
\begin{equation}\label{eq:Gamma_theta}
\Gamma = \sqrt{\frac{(3-k)E_{\rm k,iso}(\theta)}{4\pi Ac^2R^{3-k}}}
=
\left[\frac{(3-k)E_{\rm k,iso}(\theta)}{2^{5-k}\pi Ac^{5-k}t^{3-k}}\right]^{1/(8-2k)}\ ,
\end{equation}
where $R\approx 2\Gamma^2ct\propto[E_{\rm k,iso}(\theta)t]^{1/(4-k)}$ and $\rho\propto R^{-k}$. Smaller $\theta$ correspond to larger $E_{\rm k,iso}$ and therefore higher $\Gamma$ for the same $t$.
Therefore, at each $t$ there is a minimal angle $\theta = \theta_{\rm min}(t)$ for which this condition,
\begin{equation}\label{eq:theta_min}
\theta_{\rm obs}-\theta_{\rm min}(t)=\frac{1}{\Gamma[\theta_{\rm min}(t),t]}\ ,
\end{equation}
is satisfied. For all $\theta>\theta_{\rm min}(t)$ the beaming cone includes our line of sight, $\theta_{\rm obs}-\theta<1/\Gamma(\theta,t)$, i.e. their beaming cone includes our line of sight. Therefore, each such region of $\Delta\theta\sim\theta$ around $\theta$  would produce a flux corresponding to a spherical flow with the local $E_{\rm k,iso}(\theta)$ times the fraction
$f_\Omega\sim[\Gamma(\theta,t)\theta]^2$ of the solid angle $\sim\Gamma(\theta,t)^{-2}$ that would be observed for such a truly spherical flow that is actually occupied by such a region (of solid angle $\sim\theta^2$).

For PLS G, $F_\nu^{(G)}\propto f_\Omega\rho^{1/2}E_{\rm k,iso}^{(3+p)/4}t^{(3-3p)/4}$. For a given $t$, $f_\Omega\sim[\Gamma(\theta,t)\theta]^2\propto \theta^2E_{\rm k,iso}(\theta)^{1/2(4-k)}\propto\theta^{2-a/2(4-k)}$, $\rho^{1/2}\propto R^{-k/2}\propto E_{\rm k,iso}(\theta)^{-k/2(4-k)}\propto\theta^{ak/2(4-k)}$ and $E_{\rm k,iso}^{(3+p)/4}\propto\theta^{-a(3+p)/4}$ so that altogether $F_\nu^{(G)}\propto\theta^{2-a[(3+p)/4+(1-k)/2(4-k)]}$. For $k=0$ and $p=2.2$ the flux decreases with $\theta$ for $a>1.4$, which holds for the values of $a$ that are relevant for this scenario. Therefore, the observed flux is dominated by the contribution from $\theta\sim\theta_{\rm min}(t)$.

The emission becomes continuously dominated by more energetic regions of smaller $\theta_{\rm min}(t)$ so that they quickly satisfy $\theta_{\rm min}\ll\theta_{\rm obs}$ and
one can approximate $\theta_{\rm obs}-\theta_{\rm min}\approx\theta_{\rm obs}$ in Eq.~(\ref{eq:theta_min}), so that
$\Gamma[\theta_{\rm min}(t),t]\approx1/\theta_{\rm obs}=\;$constant, and therefore Eq.~(\ref{eq:Gamma_theta}) implies that $E_{\rm k,iso}[\theta_{\rm min}(t)]\propto\theta_{\rm min}(t)^{-a}\propto t^{3-k}$ and $\theta_{\rm min}(t)\propto t^{-(3-k)/a}$, which in turn imply that $f_\Omega\sim(\theta_{\rm min}/\theta_{\rm obs})^2\propto\theta_{\rm min}^2\propto t^{-2(3-k)/a}$. Given these scalings we obtain that $F_\nu^{(G)}\propto f_\Omega\rho^{1/2}E_{\rm k,iso}^{(3+p)/4}t^{(3-3p)/4}\propto t^\alpha$ for $\alpha = (3-k)(a-2)/a-k(p+1)/4$, for which an observed values of $\alpha$ could be reproduced by
\begin{equation}
    a = \frac{8(3-k)}{4(3-k)-4\alpha-k(p+1)}\;\xrightarrow{k\rightarrow0}\;
    \frac{6}{3-\alpha}\ .
\end{equation}
For the parameters relevant for GRB~170817A  ($k=0$, $p=2.2$, $\alpha=0.8$) this implies $a\approx 2.7$, which in turn implies $\theta_{\rm min}(t)\propto t^{-1.1}$, $f_\Omega\propto t^{-2.2}$ and $E_{\rm k,iso}\propto t^{3}$ (and $E_{\rm k,iso}^{(3+p)/4}\propto t^{3.9}$, $t^{(3-3p)/4}\to t^{-0.9}$) while the true energy in the region dominating the emission is $\sim\theta_{\rm min}^2E_{\rm k,iso}(\theta_{\rm min})\propto\theta_{\rm min}^{2-a}\propto t^{(3-k)(a-2)/a}\to t^{0.8}$ (for $k=0$ it always scales as $t^\alpha$).


\label{lastpage}
\end{document}